%% file: journal1_v0.9.tex
\documentclass[journal,draftclsnofoot,onecolumn,12pt]{IEEEtran}
\pdfoutput=1

\usepackage{amsthm,amssymb,graphicx,multirow,amsmath,color,amsfonts}
\usepackage[update,prepend]{epstopdf}
\usepackage[noadjust]{cite}
\usepackage[latin1]{inputenc}
\usepackage{tikz}
\usetikzlibrary{arrows,calc}		
\usepackage{bbm} 
\usepackage{pdfpages}
\usepackage{tabulary}
\usepackage{multirow}
\usepackage{comment}
\usepackage{mathtools}
\usepackage{bigints}

\include{notation}

\allowdisplaybreaks 

\begin{document}
\graphicspath{{./Figures/}}
\title{
Downlink Coverage Analysis for a Finite 3D Wireless Network of Unmanned Aerial Vehicles
}
\author{
Vishnu Vardhan Chetlur and Harpreet S. Dhillon
\thanks{The authors are with Wireless@VT, Department of ECE, Virginia Tech, Blacksburg, VA (email: \{vishnucr, hdhillon\}@vt.edu). The support of the US NSF (Grant IIS-1633363) is gratefully acknowledged.
This work was presented in part at the IEEE SPAWC, Edinburgh, UK, 2016~\cite{vishnu}. \hfill Manuscript last updated: \today.}
}
\maketitle

\begin{abstract}
In this paper, we consider a finite network of unmanned aerial vehicles (UAVs) serving a given region. Modeling this network as a uniform binomial point process (BPP), we derive the downlink coverage probability of a reference receiver located at an arbitrary position on the ground assuming Nakagami-$m$ fading for all wireless links. The reference receiver is assumed to connect to its closest transmitting node as is usually the case in cellular systems. After deriving the distribution of distances from the reference receiver to the serving and interfering nodes, we derive an exact expression for downlink coverage probability in terms of the derivative of Laplace transform of interference power distribution. In the downlink of this system, it is not unusual to encounter scenarios in which the line-of-sight (LOS) component is significantly stronger than the reflected multipath components. To emulate such scenarios, we also derive the coverage probability in the absence of fading from the results of Nakagami-$m$ fading by taking the limit $m \to \infty$. Using asymptotic expansion of incomplete gamma function, we concretely show that this limit reduces to a redundant condition. Consequently, we derive an accurate coverage probability approximation for this case using dominant interferer-based approach in which the effect of dominant interferer is exactly captured and the residual interference from other interferers is carefully approximated. We then derive the bounds of the approximate coverage probability using Berry-Esseen theorem. Our analyses reveal several useful trends in coverage probability as a function of height of the transmitting nodes and the location of reference receiver on the ground.
\end{abstract}

\section{Introduction} \label{sec:intro}

With significant advancements in the drone technology, like increased payload capacity, longer average flight time, better power management techniques, and the capability to harvest solar energy, unmanned aerial vehicles (UAVs) can serve a multitude of purposes such as surveillance, localization and communication, making them a flexible solution to augment and enhance the capabilities of the current cellular systems. They provide an especially attractive solution to provide connectivity in the wake of disasters and accidents, which may completely cripple the terrestrial networks due to damaged equipment and/or loss of power~\cite{uav_safety,gen1}. In general, UAVs provide a realistic solution in scenarios where there is a temporary need for network resources. These could include first responder situations, such as the one discussed above, or even usual civilian scenarios, such as football games or concerts. In order to provide short-term connectivity in such scenarios, temporary deployment of UAVs may be faster and more cost-effective compared to the temporary installation of conventional base stations. They are also currently being investigated as a possible candidate for providing ubiquitous connectivity in remote areas that lack traditional cellular infrastructure. While there is no doubt about the deployment flexibility and general benefits of UAVs, their performance in terms of the coverage and capacity provided to the terrestrial users is not quite well understood. This is especially true for a realistic use case of finite UAV networks, where we have a given number of UAVs serving users in a given region (such as a city). In this paper, we use tools from stochastic geometry to derive downlink signal-to-interference ratio distribution for this setup, which immediately provides useful insights into the coverage performance of the resulting three-dimensional network. Several intermediate results derived in this paper are also of general interest for the analysis of finite wireless networks.

\subsection{Motivation and Related Work} 
The improvements in payload capacity and prolonged flight times have enabled the commercial use of UAVs, especially for communication purposes. UAV networks differ significantly from conventional wireless networks in terms of the mobility, energy constraints, as well as the propagation conditions. This has stimulated interest in the design of application-oriented protocols for the effective utilization of aerial networks~\cite{sur1,sur2,sur3}. For instance, a cluster-based protocol, which improves the resilience to frequent link failures resulting from the motion of UAVs, has been proposed in~\cite{sur2}. The flexibility offered by the mobility of UAVs has motivated a lot of {\em algorithmic} research efforts towards finding efficient trajectories and deployment strategies aimed at optimizing different network resources~\cite{belllabs,optuplink,optimallap,wirelessRelay,mohmdl,halim,UAVpath}. For instance, an algorithm to optimize transmit power and frequency spectrum for autonomous self-deployment was proposed in~\cite{belllabs}. An adaptive algorithm for adjusting the UAV heading was proposed in~\cite{optuplink} to improve the uplink performance and minimize mutual interference. An approach to optimize the altitude of UAVs to maximize coverage on the ground was proposed in~\cite{optimallap}. The performance of UAVs acting as relays between terrestrial users and base stations was investigated in~\cite{wirelessRelay}. The problem of efficient placement of UAVs with slightly different objectives was studied in~\cite{mohmdl,halim}.


Another direction of research, which is somewhat complementary to the one discussed above, is to develop techniques for the realistic system-level analysis of UAV networks. As is the case in terrestrial networks, such as cellular networks, these techniques can then be used to compare the performance of different deployment strategies and to benchmark their performance against standard baselines. In the case of UAV networks, the system-level performance has mostly been studied through field tests and simulations~\cite{ft1,ft2,ft3}. For instance, in~\cite{ft2} the outage time and average goodput were compared for different routing algorithms using real-world experiments. While field-tests or simulations can provide initial insights into the behavior of the network, these methods are usually not scalable when the number of simulation parameters is large. One way of reducing the dimensionality of such problems is to endow the locations of the nodes with a distribution, which additionally allows the use of powerful tools from stochastic geometry to derive easy-to-use expressions for key performance metrics. While stochastic geometry has already emerged as a preferred tool for the analysis of ad hoc and cellular networks~\cite{tutorial_jeff}, its potential has not yet been exploited for the analysis of UAV networks. One relevant prior art is~\cite{uavd2d}, which studies the co-existence of a device-to-device (D2D) communication network with a single UAV. In this work, we will develop the first comprehensive model aimed at the downlink analysis of a finite multi-UAV network using tools from stochastic geometry. 

While infinite homogeneous Poisson Point Process (PPP) has become a canonical model for the spatial locations of terrestrial base stations~\cite{tutorial_jeff}, it is not quite suitable for UAV networks, especially when a given number (likely small) of UAVs is deployed to cover a given finite region. For such scenarios, a simple yet reasonable model for the spatial distribution of UAVs is the homogeneous binomial point process (BPP)~\cite{haenggi2013stochastic,distbpp}. More sophisticated models incorporating inter-point interaction are usually far less tractable. While BPP has not yet been used for the analysis of UAVs, it has received significant attention for the analysis of terrestrial networks with a given number of nodes. Until recently, however, the analysis was focused on ad hoc networks, in which a given number of nodes were assumed to be distributed uniformly at random in a circular region with the reference receiver located at the center of the circle~\cite{zhang,venugopal,gong,valenti,durrani}. The outage probability for this reference receiver is then derived assuming that it is served by a reference transmitter located at a fixed distance (not a part of the BPP). In order to model cellular systems meaningfully with this setup, two key generalizations are required: (i) reference receiver can lie anywhere in the region, and (ii) serving base station for the reference receiver will be chosen from the BPP itself. For the latter, it is reasonable to assume that the reference receiver is served by the closest base station from the BPP. The exact analysis of this finite cellular network setup was done very recently in~\cite{mehrnazbpp}. An approximate analysis of a related setup also appears in~\cite{BanEckJ2015}. Building on the distance distributions derived in~\cite{mehrnazbpp,khaDurJ2013}, we will perform downlink analysis for an arbitrarily located user on the ground that is served by a finite network of UAVs. In addition to providing the first such system-level analysis of a finite UAV network, several intermediate results provide constructs that are more generally applicable to the analysis of finite wireless networks. With this brief discussion, we now provide a precise summary of our contributions.%

\subsection{Contributions}
{\em{Modeling of finite three-dimensional network.}} We develop a general framework for the downlink coverage analysis of finite three-dimensional networks under a fairly general channel fading model. In particular, we consider a finite network of a given number of UAVs whose locations are modeled as a uniform BPP in a plane at a fixed altitude above the ground. The reference receiver is assumed to be located at some arbitrary position on the ground. As noted above, we assume that the reference receiver connects to its closest transmitting UAV node. We then derive the distribution of distances from the reference receiver to the serving and interfering nodes. As the air-to-ground channel models are still under active investigation, we assume Nakagami-$m$ fading, which allows us to control the severity of multi-path fading (to emulate variety of scenarios) while retaining analytical tractability. In some deployment scenarios, the LOS component may be significantly stronger than the reflected multipath components, which means it may be reasonable to study system performance in the absence of small-scale fading. Such scenarios, which will henceforth be referred to as ``no-fading'' environments, are also studied in detail.

{\em{Coverage probability.}} We derive an exact expression for coverage probability of the reference receiver located at some arbitrary position on the ground for Nakagami-$m$ fading channel in terms of the Laplace transform of interference power distribution. While this problem has been studied in~\cite{durrani} for terrestrial networks, the complexity and form of the final expression forbids any further analysis or simplifications. In this paper, we develop an approach that explores the possibility of deriving the results for no-fading case from the results of Nakagami-$m$ fading by applying the limit $m\to \infty$. We use well-established mathematical results, especially asymptotic expansion of incomplete gamma function, in the evaluation of this non-trivial limit. Quite interestingly, we discover that this limit renders a redundant condition for coverage probability, thereby not yielding an explicit expression for the no-fading case. As a result, we derive a simple yet accurate approximation for coverage probability using dominant interferer-based approach in which the effect of dominant interferer is accurately captured and the aggregate interference from the rest of the interferers is carefully approximated. We then derive the bounds of the approximate coverage probability using Berry-Esseen theorem (BET)~\cite{BET}. 

{\em{Performance analysis.}} We analyze the trends in coverage probability for different system parameters such as the altitude of the UAVs, the location of the reference receiver, the path-loss exponent of the channel, and the Nakagami-$m$ fading parameter. We demonstrate that the coverage probability of the reference receiver degrades as the altitude of the transmitting nodes increases when the area over which the transmitters are scattered remains unchanged. We also observe that the coverage probability of the reference receiver increases as the path-loss exponent of the channel increases. These observations offer useful guidelines for the system design. Several intermediate mathematical results are of general interest to the analysis of finite networks.
 
\section{System Model} \label{sec:SysMod}
\begin{figure}
\centering
\begin{minipage}{.5\textwidth}
\centering
\includegraphics[scale=0.3]{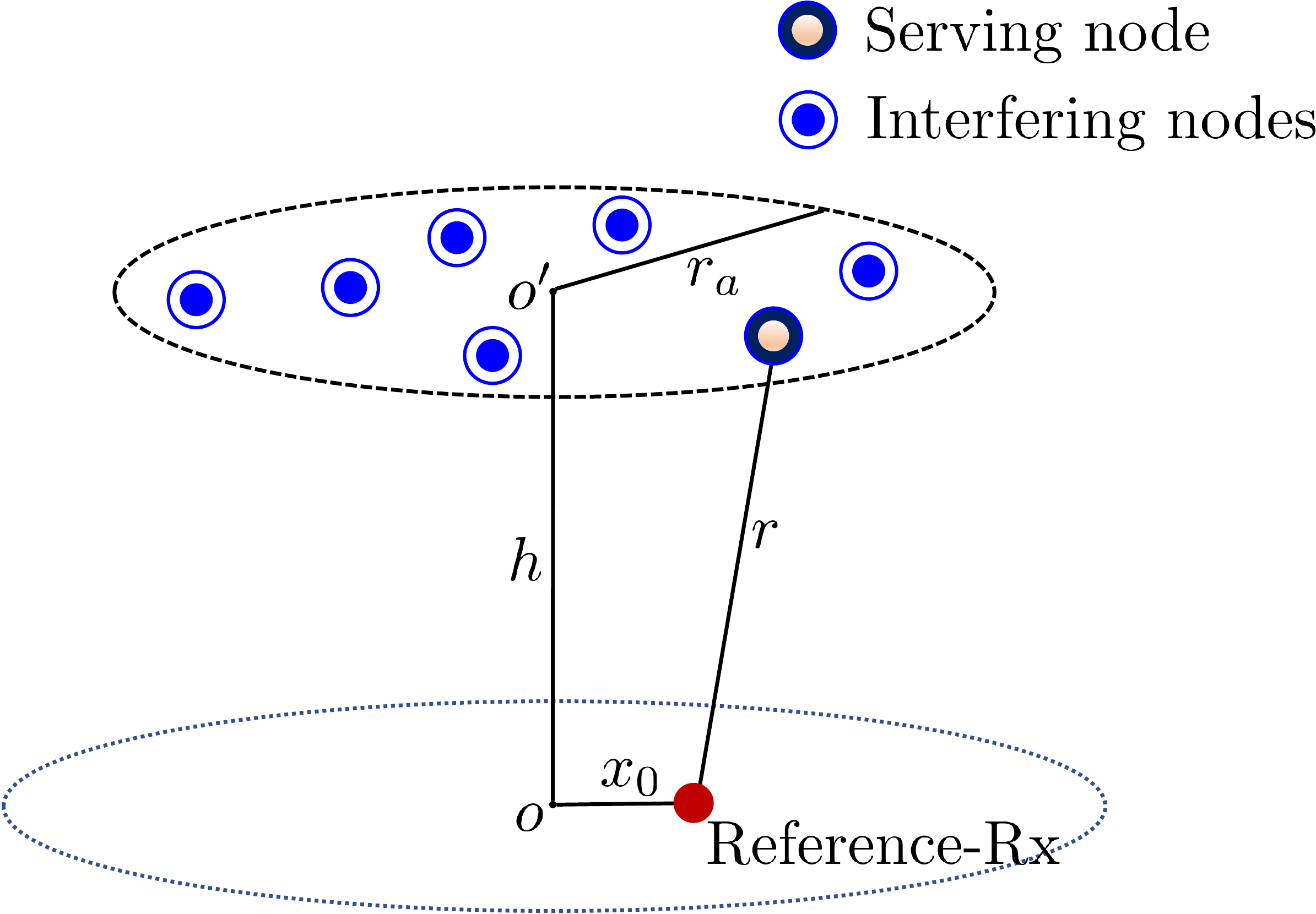}
\caption{Illustration of the system model}
\label{fig:sysmod1}
\end{minipage}%
\begin{minipage}{.5\textwidth}
\centering
\includegraphics[scale=0.3]{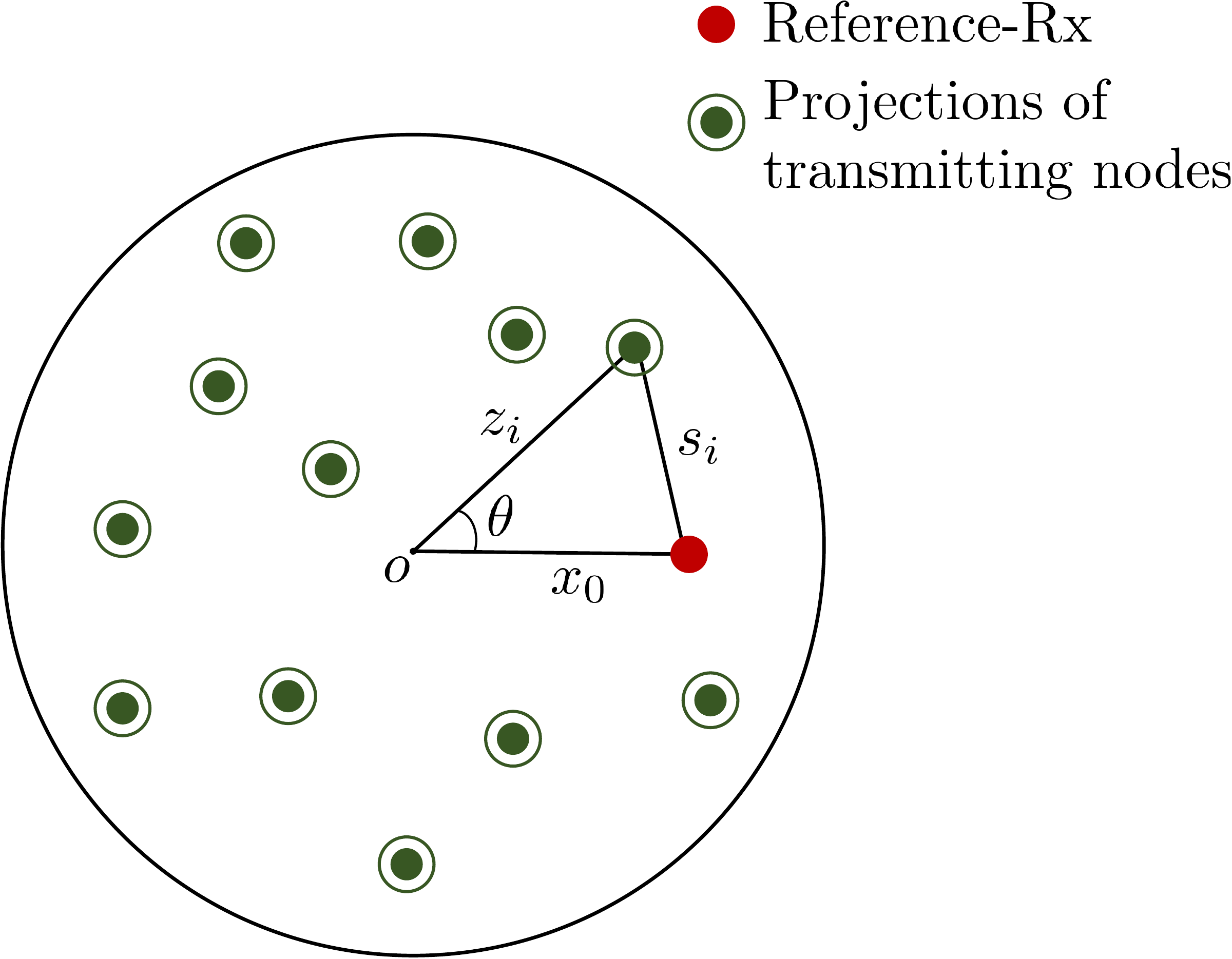}
\caption{Illustration of projection of location of the nodes onto the ground plane.}
\label{fig:sysmod2}
\end{minipage}
\end{figure}

We consider a network of $N$ transmitting devices (UAVs) uniformly distributed in a finite area forming a BPP. While the devices can be strategically placed to optimize the network utility, in the absence of exact traffic patterns, these optimal locations are not known, which justifies the BPP assumption. As discussed in the previous section, BPP is a finite-network analogue of a PPP, which is a popular model for infinite networks. This is a simple yet reasonable first step towards comprehensive understanding of these networks. The locations of the devices are uniformly distributed in a disk $b(o',r_a)$ of radius $r_a$ centered at $o' = (0,0,h)$, as depicted in Fig. \ref{fig:sysmod1}. For simplicity, we assume that all the devices are positioned at the same height $h$. The locations of the devices are denoted by $\{{\bf{y}}_i\}_{i=1:N} \equiv \Phi \subset \nbbR^2$. The distance of the $i^{th}$ node from $o'$ is denoted by $Z_i = \|{\bf{y}}_i - o'\|$ for $1 \leq i \leq N$. The sequence of distances $\{Z_i\}$ is unordered, which means that the indices are assigned arbitrarily to the nodes.
For this setup, we perform downlink coverage analysis for a reference receiver located at some arbitrary position on the ground, at a distance of $x_0$ from the origin $o \equiv (0,0,0)$. For brevity, the {\em reference receiver} will henceforth be referred to as only the {\em receiver}. Since the point process is invariant to the orientation of the axes, we can assume, without loss of generality, that the receiver lies on the $x$-axis, i.e., the location of the receiver is ${\bf{x}} \equiv (x_0,0,0)$. The distance between the receiver and the projection of the location of $i^{th}$ transmitting device onto the ground plane is denoted by $S_i$, as shown in Fig. \ref{fig:sysmod2}. The angular separation between the receiver position and the projection of the device location is denoted by $\theta$, which is uniformly distributed in the range $[0,\  2\pi)$. Note that $S_i, Z_i, x_0,$ and $\theta$ are related by the cosine rule. The receiver is assumed to connect to its closest transmitter from $\Phi$. The unordered set of distances from the receiver to the transmitters is denoted by $\{W_i\} = \{\sqrt{S_i^2 + h^2}\} $. The ordered set of distances is denoted by $\{W_{(i)}\}_{i=1:N}$, where $W_{(i)}$ is the distance between the receiver and the $i^{th}$ closest transmitter to the receiver. From this set, the serving distance is denoted by $R=W_{(1)}$. The distance to the closest interfering node is denoted by $U_1 = W_{(2)}$ and the unordered set of distances between the receiver and remaining $N-2$ interferers is denoted by $\{U_i\}_{i=2:N-1}$. 

We assume that all the nodes transmit at the same power. For notational simplicity, we assume that the thermal noise is negligible as compared to the interference experienced at the receiver and is hence ignored. Owing to the limited knowledge of air-to-ground channel models for low altitude platforms, we choose Nakagami-$m$ fading, which is a generalized model that mimics various fading environments. We denote the Nakagami-$m$ fading parameter for the serving link and interfering links by $m_0$ and $m$, respectively. We restrict the values of $m_0$ to integers for analytical tractability. The signal-to-interference ratio (SIR) at the receiver is 
\begin{equation*}
\sir = \frac{G_0 R^{-\alpha}}{\sum_{i=1}^{N-1} G_i U_i^{-\alpha}},
\end{equation*}
where $\alpha > 2$ is the path-loss exponent, and $G_0$ and $G_i$ are channel fading gains corresponding to serving and interfering links. The channel gains $G_0$ and $G_i$ follow a gamma distribution with the probability density function (PDF)~\cite{wackerly}
\begin{align*}
f_G(g) = \frac{m^m g^{m-1}}{\Gamma(m)}\exp({-mg}).
\end{align*}

\section{Coverage Probability}\label{sec:pc}
In this section, we derive the coverage probability for the setup introduced in the previous section. Using properties of gamma function, we will express it in terms of the derivative of Laplace transform of interference power distribution. We also attempt to obtain the coverage probability for no-fading channel from the results of Nakagami-$m$ fading by taking the limits $m \to \infty$ and $m_0 \to \infty$. Quite interestingly, we discover that this limit renders a redundant condition for coverage probability, thereby not yielding an explicit expression for the no-fading case. Hence, we provide an alternate approach to compute the coverage probability in this case by approximating the aggregate interference from devices other than the dominant interferer with a normal random variable and capturing the effect of dominant interferer exactly. 
\subsection{Relevant Distance Distributions}
We begin our analysis by characterizing the distribution of distances between the receiver and the transmitters in this subsection. While somewhat similar expositions about distance distributions can be found in~\cite{mehrnazbpp,khaDurJ2013} in the context of terrestrial networks, the distributions corresponding to the dominant interferer-based approach are unique to this paper because this paper is the first one to apply that approach to the analysis of finite cellular networks.

\begin{lemma}\label{lem:cdfwi}
The distances from the receiver to the set of independently and uniformly distributed transmitting devices, denoted by $\{W_i\}$, conditioned on $x_0 = \|{\bf{x}}\|$, are independent and identically distributed (i.i.d.), with the cumulative distribution function (CDF) of each element given by 
\begin{align}
F_{W_i}(w_i | x_0) =  
\begin{cases}
	F_{W_{i,1}}(w_i | x_0 ), &     h \leq w_i \leq w_m \\
	F_{W_{i,2}}(w_i | x_0 ), &     w_m < w_i \leq w_p
\end{cases}
,
\end{align}
with 
\begin{align}
&F_{W_{i,1}}(w_i | x_0 ) = \frac{w_i^2-h^2}{r_a^2}, \quad F_{W_{i,2}}(w_i | x_0 ) = \frac{w_i^2-h^2}{\pi r_a^2}\left(\theta^* - \frac{1}{2}\sin 2\theta^*\right) + \frac{1}{\pi}\left(\phi^* - \frac{1}{2}\sin 2\phi^*\right), 
\end{align}
where
\begin{align}\notag
&\theta^* = \arccos\bigg(\frac{w_i^2+x_0^2-d^2}{2 x_0 \sqrt{w_i^2-h^2} }\bigg),\ \phi^* = \arccos\bigg(\frac{x_0^2+d^2-w_i^2}{2 x_0 r_a}\bigg), \\ \notag &w_m = \sqrt{(r_a-x_0)^2 + h^2},\ w_p = \sqrt{(r_a+x_0)^2 + h^2}, \text{ and } d = \sqrt{r_a^2 + h^2}. 
\end{align} 
\end{lemma}
\begin{IEEEproof}
See Appendix \ref{app:cdfwi}. 
\end{IEEEproof}

\begin{lemma}\label{lem:pdfwi}
The PDF of $W_i$ conditioned on $x_0$ is  
\begin{align}
f_{W_i}(w_i | x_0) =  
\begin{cases}
	f_{W_{i,1}}(w_i | x_0 ), &     h \leq w_i \leq w_m \\
	f_{W_{i,2}}(w_i | x_0 ), &     w_m < w_i \leq w_p
\end{cases}
,
\end{align}
with 
\begin{align}
&f_{W_{i,1}}(w_i | x_0 ) = \frac{2 w_i}{r_a^2}, \quad f_{W_{i,2}}(w_i | x_0 ) = \frac{2 w_i}{\pi r_a^2}\arccos\bigg(\frac{w_i^2+x_0^2-d^2}{2 x_0 \sqrt{w_i^2-h^2} }\bigg), 
\end{align}
where $\ w_m = \sqrt{{s_m}^2 + h^2},\ w_p = \sqrt{{s_p}^2 + h^2},$ \text{and} $\ d = \sqrt{r_a^2+h^2}.$
\end{lemma}
\begin{IEEEproof}
$f_{W_i}(w_i|x_0)$ can be derived by taking the derivative of $F_{W_i}(w_i|x_0)$ from Lemma \ref{lem:cdfwi} with respect to $w_i$.
\end{IEEEproof}

For a receiver located at the origin $o$, this piece-wise expression for the PDF reduces to a simple expression, which is given in the following Corollary.
\begin{cor}
The set of distances from a receiver located at the origin to the transmitting devices are i.i.d. with the PDF of each element given by 
\begin{align}
f_{W_i}(w_i) = \begin{dcases}
\frac{2w_i}{r_a^2}, \qquad     &h \leq w_i \leq d \\
0, \qquad &otherwise
\end{dcases}.
\end{align}
\end{cor}

\begin{lemma}\label{lem:pdfu1}
The PDF of the serving distance $R$ conditioned on $x_0$ is  
\begin{align}\label{equ:pdfu1}
f_{R}(r | x_0) =  
\begin{cases}
	f_{R,1}(r | x_0 ), &     h \leq r \leq w_m \\
	f_{R,2}(r | x_0 ), &     w_m < r \leq w_p
\end{cases}
,
\end{align}
with 
\begin{align}
&f_{R,1}(r | x_0 ) =  N\big( 1 - F_{W_{i,1}}(r|x_0) \big)^{N-1}f_{W_{i,1}}(r|x_0), \\
&f_{R,2}(r | x_0 ) =  N\big( 1 - F_{W_{i,2}}(r|x_0) \big)^{N-1}f_{W_{i,2}}(r|x_0). 
\end{align}
\end{lemma}
\begin{IEEEproof}
See Appendix \ref{app:pdfu1}.
\end{IEEEproof}

If the receiver is located at the origin, the above result reduces to a simple expression, which is given in the next Corollary.
\begin{cor}
The PDF of the serving distance $R$ for a receiver located at the origin is
\begin{align}
f_{R}(r) =  N\Bigg(\frac{2r}{r_a^2}\Bigg)\Bigg(\frac{d^2-r^2}{r_a^2}\Bigg)^{N-1},\qquad     h \leq r \leq d .
\end{align}
\end{cor}
\begin{figure}
\centering
\includegraphics[scale=.32]{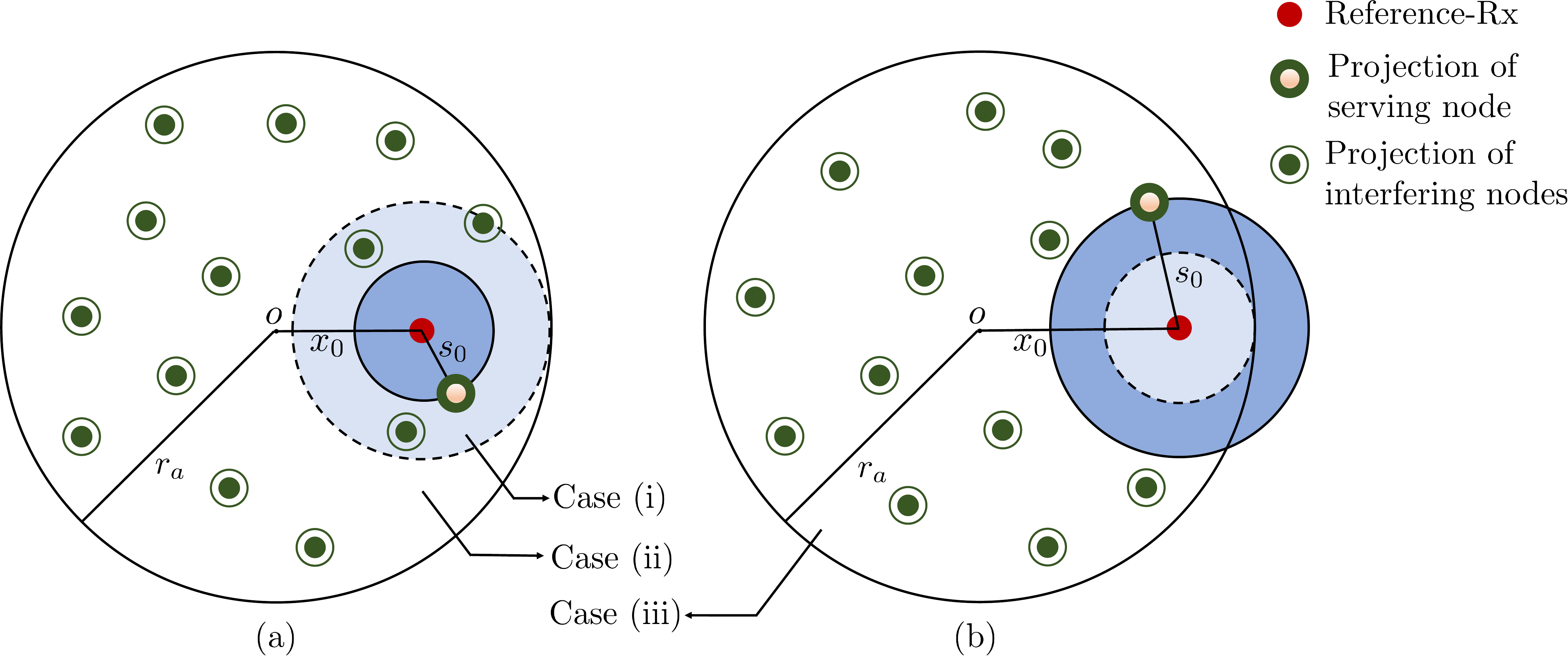}
\caption{System model of the cases: (a) $s_0<r_a-x_0$, and (b) $s_0>r_a-x_0$.}
\label{fig:cases}
\end{figure}
We now derive the distribution of distance between the receiver and interferers, conditioned on the serving distance $R$. This distribution will be useful in characterizing the interference experienced at the receiver.
\begin{lemma}\label{lem:pdfuigu1}
Conditioned on the serving distance $R$, the unordered set of distances between the interferers and the receiver, $\{U_i\}_{i=1:N-1}$, are i.i.d. with the PDF of each element given by
\begin{align}
f_{U_i}(u_i | r, x_0) = 
\begin{dcases}
\frac{f_{W_{i,k}}(u_i|x_0)}{1-F_{W_{i,l}}(r|x_0)}, & r \leq u_i \leq w_p \\
0,   & otherwise
\end{dcases},
\end{align}
with
\begin{alignat*}{4}\notag
k\, = 1, \, l\, = 1,&& \qquad  h &\leq r \leq w_m,\  &r &\leq  u_i \leq w_m \\ \notag
k\, = 2, \, l\, = 1,&& \qquad  h &\leq r \leq w_m,\ &w_m &\leq  u_i \leq w_p \\ \notag
k\, = 2, \, l\, = 2,&& \qquad  w_m &\leq r \leq w_p,\  &r &\leq  u_i \leq w_p .
\end{alignat*}
\end{lemma}
\begin{IEEEproof}
See Appendix \ref{app:pdfuigu1}.
\end{IEEEproof}

For the case where the receiver is located at the origin, the conditional PDF of distances between the receiver and the interferers is given in the following Corollary.
\begin{cor}
For a receiver situated at the origin, the PDF of distances between the receiver and the interferers $U_i$ conditioned on the serving distance $R$ is given by
\begin{align}
f_{U_i}(u_i | r) =\begin{dcases}
\frac{2u_i}{d^2 - r^2}, \qquad &r \leq u_i \leq d \\
0 \qquad &otherwise
\end{dcases}.
\end{align}
\end{cor}
We now characterize the joint distribution of serving distance $R$ and the distance between the receiver and the dominant interferer $U_1$. This distribution holds the key to the derivation of approximate coverage probability using the dominant interferer approach (for the no-fading case) as discussed in the sequel. Note that the dominant interferer in the absence of fading is the second closest transmitter to the receiver.
Since it is easy to visualize and understand these cases in two dimensions, we consider the projections of the transmitting nodes onto the ground plane. We will then use these results to derive the distributions in the actual three-dimensional setup. 
Let the distance from the receiver to the projections of locations of serving transmitter and dominant interferer be denoted by $S_0$ and $S_1$, respectively. Using the same argument as presented in Lemma \ref{lem:pdfuigu1}, the piece-wise nature of joint distribution of $R$ and $U_1$ can be attributed to the following three cases: (i) both the disks $b({\bf{x}},s_0)$ and $b({\bf{x}},s_1)$ are entirely contained in the disk $b(o,r_a)$, (ii) the disk $b({\bf{x}},s_0)$ is entirely contained in $b(o,r_a)$, while $b({\bf{x}},s_1)$ overlaps only partially with $b(o,r_a)$, and (iii) both the disks $b({\bf{x}},s_0)$ and $b({\bf{x}},s_1)$ overlap only partially with $b(o,r_a)$. Note that the converse of case (ii) is not valid since $s_1 > s_0$.

\begin{lemma}\label{lem:jpdfu1u2}
The conditional joint PDF of serving distance $R$ and the distance of the dominant interferer from the receiver $U_1$ is given by
\begin{align}
& f_{R,U_1}(r,u_1 | x_0) = 
&\begin{dcases}
N(N-1)f_{W_{i,k}}(r|x_0)f_{W_{i,l}}(u_1|x_0) \\ 
\qquad \times [1-F_{W_{i,l}}(u_1 | x_0)]^{N-2},  & h \leq r \leq  u_1 \leq w_p \\
0          & otherwise
\end{dcases},
\end{align}
with
\begin{alignat*}{4}\notag
k\, = 1, \, l\, = 1,&& \qquad  h &\leq r \leq w_m,\  &r &\leq  u_1 \leq w_m \\ \notag
k\, = 2, \, l\, = 1,&& \qquad  h &\leq r \leq w_m,\ &w_m &\leq  u_1 \leq w_p \\ \notag
k\, = 2, \, l\, = 2,&& \qquad  w_m &\leq r \leq w_p,\  &r &\leq  u_1 \leq w_p. 
\end{alignat*}
\end{lemma}
\begin{IEEEproof}
See Appendix \ref{app:jpdfu1u2}. 
\end{IEEEproof}
For the case where the receiver is at the origin, this result reduces to a simple expression, which is given in the next Corollary. 
\begin{cor}
For a receiver located at the origin, the joint PDF of the serving distance $R$ and the distance of the dominant interferer from the receiver $U_1$ is 
\begin{align}
f_{R,U_1}(r,u_1) = \begin{dcases}
\frac{N(N-1)(d^2-u_1^2)^{N-2}4ru_1}{(d^2-h^2)^N}, \qquad &h \leq r < u_1 \leq d \\
0 \qquad &otherwise
\end{dcases}.
\end{align}
\end{cor}
We now determine the distribution of the set of distances between the receiver and the interferers $\{U_i\}_{i=2:N-1}$, conditioned on the serving distance $R$, and the distance of the dominant interferer from the receiver $U_1$. Let the distances from the receiver to the projections of remaining $N-2$ interferers be denoted by $\{S_i\}_{i=2:N-1}$. The piece-wise distribution of these distances depends on the following four cases: (i) all the disks $b({\bf{x}},s_0)$,  $b({\bf{x}},s_1)$, and  $b({\bf{x}},s_i)$ are completely contained in  $b(o,r_a)$, (ii) the disks $b({\bf{x}},s_0)$ and  $b({\bf{x}},s_1)$  are completely contained in  $b(o,r_a)$, while the disks $b({\bf{x}},s_i)$ overlap only partially with $b(o,r_a)$, (iii) the disk  $b({\bf{x}},s_0)$ is completely contained in the disk $b(o,r_a)$, while $b({\bf{x}},s_1)$ and $b({\bf{x}},s_i)$ partially overlap with $b(o,r_a)$, and (iv) all the disks $b({\bf{x}},s_0)$, $b({\bf{x}},s_1)$, and $b({\bf{x}},s_i)$ overlap partially with $b(o,r_a)$. Note that other combinations are not valid because $s_0 < s_1  < s_i$.

\begin{lemma}\label{lem:pdfui}
Conditioned on the serving distance $R$ and the distance of the dominant interferer from the receiver $U_1$, the set of distances between the remaining interferers and the receiver, $\{U_i\}_{i=2:N-1}$, are i.i.d. with the PDF of each element given by
\begin{align}
f_{U_i}(u_i | r, u_1, x_0) = 
\begin{dcases}
\frac{f_{W_{i,k}}(u_i|x_0)}{1-F_{W_{i,l}}(u_1|x_0)}, & u_1 \leq u_i \leq w_p \\
0,   & otherwise
\end{dcases},
\end{align}
with 
\begin{alignat*}{5}\notag
k\, = 1, \, l\, = 1,&& \   h &\leq r \leq w_m,\  &r &\leq  u_1 \leq w_m,\  &u_1 &\leq  u_i \leq w_m \\ \notag
k\, = 2, \, l\, = 1,&& \   h &\leq r \leq w_m,\  &r &\leq  u_1 \leq w_m,\  &w_m &\leq  u_i \leq w_p \\ \notag
k\, = 2, \, l\, = 2,&& \   h &\leq r \leq w_m,\ &w_m &\leq  u_1 \leq w_p,\  &u_1 &\leq  u_i \leq w_p \\ \notag
k\, = 2, \, l\, = 2,&& \   w_m &\leq r \leq w_p,\  &r &\leq  u_1 \leq w_p,\  &u_1 &\leq  u_i \leq w_p . 
\end{alignat*}
\normalsize
\end{lemma}
\begin{IEEEproof} 
The proof follows along the same lines as that of Lemma \ref{lem:pdfuigu1}.
\end{IEEEproof}
The following Corollary gives the conditional PDF of distances of interferers from the receiver $U_i$ for the special case where the receiver is located at the origin.
\begin{cor}
Conditioned on the serving distance $R$ and the distance of the dominant interferer from the receiver $U_1$, the set of distances between the receiver and the interferers, $\{U_i\}_{i=2:N-1}$, are i.i.d. with the PDF of each element given by
\begin{align}
f_{U_i}(u_i | r, u_1) = \begin{dcases}
\frac{2u_i}{d^2 - u_1^2}, \qquad &u_1 \leq u_i \leq d \\
0 \qquad &otherwise
\end{dcases}.
\end{align}
\end{cor}
Now that we have determined the necessary distance distributions, we will proceed to derive the coverage probability for the receiver under Nakagami-$m$ fading in the following subsection.

\subsection{Coverage Probability under Nakagami-$m$ Fading Channels}
The coverage probability is formally defined as the probability with which the SIR at the receiver exceeds a pre-determined threshold necessary for a successful communication. Recall that the SIR at the receiver in Nakagami-$m$ fading is given by
\begin{align*}
\sir = \frac{G_0 R^{-\alpha}}{\sum_{i=1}^{N-1} G_i U_i^{-\alpha}},
\end{align*}
where $\alpha > 2$ is the path-loss exponent, and $G_0$ and $G_i$ are channel fading gains with parameters $m_0$ and $m$, respectively. We denote the set of gains for the interfering links by $\mathcal{G}=\{G_i\}$ and the set of distances of the interferers from the receiver by $\mathcal{U}=\{U_i\}$. We first calculate the coverage probability, conditioned on $R$, as a derivative of conditional Laplace transform of interference power distribution, which is given in the following Lemma. 

\begin{lemma}\label{lem:LI}
The Laplace transform of interference power distribution conditioned on the serving distance $R$ is
\begin{align}
\calL_I(s| r, x_0) = \begin{dcases}
\calA(s,r,x_0) \quad h \leq r \leq w_m\\
\calB(s,r,x_0) \quad w_m \leq r \leq w_p
\end{dcases},
\end{align}
where 
\begin{align}\notag
\calA(s,r,x_0) &= \Bigg[ \bigintsss_r^{w_m} \bigg( 1 + \frac{s u_i^{-\alpha}}{m} \bigg)^{-m} \frac{f_{W_{i,1}}(u_i|x_0)}{1-F_{W_{i,1}}(r|x_0)} {\rm d}u_i \\
& \qquad \qquad \qquad + \bigintsss_{w_m}^{w_p} \bigg( 1 + \frac{s u_i^{-\alpha}}{m} \bigg)^{-m} \frac{f_{W_{i,2}}(u_i|x_0)}{1-F_{W_{i,1}}(r|x_0)} {\rm d}u_i \Bigg]^{N-1}, \\
\calB(s,r,x_0) &= \Bigg[ \bigintsss_r^{w_p} \bigg( 1 + \frac{s u_i^{-\alpha}}{m} \bigg)^{-m} \frac{f_{W_{i,2}}(u_i|x_0)}{1-F_{W_{i,2}}(r|x_0)} {\rm d}u_i \Bigg]^{N-1}.
\end{align}
\end{lemma}
\begin{IEEEproof}
See Appendix \ref{app:LI}. 
\end{IEEEproof}
For the case where the receiver is located at the origin, this result reduces to a simple expression, which is given in the next Corollary.
\begin{cor}
For a receiver located at the origin, the Laplace transform of the interference power distribution conditioned on the serving distance $R$ is
\begin{align}
\calL_I(s|r) = \bigintssss_r^d \bigg(1 + \frac{su_i^{-\alpha}}{m} \bigg)^{-m} \frac{2 u_i}{d^2-r^2} {\rm d}u_i
\end{align}
\end{cor}
Using the Laplace transform of conditional interference power distribution, we derive the coverage probability in the following theorem.
\begin{thm}\label{thm:pcm}
The coverage probability of the receiver in Nakagami-$m$ fading channel is 
\begin{align} \notag
&\pc = \bigintss_h^{w_m} \Bigg( \sum_{k=0}^{m_0-1} \frac{(-1)^k}{k!} \bigg[ \frac{\partial^k}{\partial s^k} \calA(s,r,x_0) \bigg]_{s=m_0 \T r^{\alpha}} \Bigg) N\big( 1 - F_{W_{i,1}}(r|x_0) \big)^{N-1}f_{W_{i,1}}(r|x_0){\rm d}r \\
&+ \ \bigintss_{w_m}^{w_p} \Bigg( \sum_{k=0}^{m_0-1} \frac{(-1)^k}{k!} \bigg[ \frac{\partial^k}{\partial s^k} \calB(s,r,x_0) \bigg]_{s=m_0 \T r^{\alpha}} \Bigg) N\big( 1 - F_{W_{i,2}}(r|x_0) \big)^{N-1}f_{W_{i,2}}(r|x_0){\rm d}r. 
\end{align}
\end{thm}
\begin{IEEEproof}
We first derive the conditional coverage probability as follows:
\begin{align} \notag
\P\big( \sir > \T | R, x_0\big) &= \nbbE_I \bigg[\P\Big( G_0 > \T R^{\alpha} I | R, I, x_0 \Big) \bigg] \\ \notag
&\stackrel{(a)}{=} \nbbE_I \Bigg[ \frac{\Gamma\big(m_0, m_0 \T r^{\alpha} I \big)}{\Gamma(m_0)}\bigg| R, x_0 \Bigg]\\ \notag
&\stackrel{(b)}{=} \nbbE_I \Bigg[ \sum_{k=0}^{m_0-1} \frac{\Big(m_0 \T r^{\alpha} I \Big)^k}{k!} \exp\big(-m_0 \T r^{\alpha} I\big) \bigg| R, x_0\Bigg] \\ \label{equ:pcm2}
&= \sum_{k=0}^{m_0-1} \frac{(-m_0 \T r^{\alpha})^k}{k!} \bigg[ \frac{\partial^k}{\partial s^k} \calL_I(s| r, x_0) \bigg]_{s=m_0 \T r^{\alpha}},
\end{align}
where (a) follows from the CCDF of gamma random variable $G_0$, and (b) follows from the definition of incomplete gamma function for integer values of $m_0$. The overall coverage probability can now be obtained by substituting the Laplace transform of interference distribution from Lemma \ref{lem:LI} in \eqref{equ:pcm2} and deconditioning the resulting expression over $R$. This completes the proof.
\end{IEEEproof}

\subsection{Limiting Case of No-fading
}
In this subsection, we attempt to derive the coverage probability for a no-fading environment from the results of Nakagami-$m$ fading by applying the limits $m \to \infty$ and $m_0 \to \infty$. This approach has mostly been overlooked in the literature due to the complexity of the results of Nakagami-$m$ fading. A partial attempt to compute this limit was made in~\cite{ralph} where the limit $m \to\infty$ is applied only on the interfering links but not on the desired link. We show that it is more challenging to evaluate the limit for the desired link $m_0 \to \infty$. Our analysis will rely on an asymptotic expansion of incomplete gamma function.

From our approach, we observe that it is convenient to take this limit in one of the intermediate steps in the derivation of $\pc$ under Nakagami-$m$ fading. Substituting \eqref{equ:pcm1} in \eqref{equ:pcm2}, we get the conditional coverage probability in Nakagami-$m$ fading as follows:
\begin{align} \notag
\P\big( \sir > \T | R, x_0\big) &= \sum_{k=0}^{m_0-1} \frac{(-m_0 \T r^{\alpha})^k}{k!} \bigg[ \frac{\partial^k}{\partial s^k} \nbbE_{\ncalU} \bigg[ \prod_{i=1}^{N-1} \bigg(1 + \frac{s U_i^{-\alpha}}{m} \bigg)^{-m} \bigg| R, x_0\bigg] \bigg]_{s=m_0 \T r^{\alpha}} \\
&\stackrel{(a)}{=} \nbbE_{\ncalU} \Bigg(\sum_{k=0}^{m_0-1} \frac{(-m_0 \T r^{\alpha})^k}{k!} \bigg[ \frac{\partial^k}{\partial s^k} \bigg[ \prod_{i=1}^{N-1} \bigg(1 + \frac{s U_i^{-\alpha}}{m} \bigg)^{-m} \bigg] \bigg]_{s=m_0 \T r^{\alpha}} \bigg| R, x_0 \Bigg),
\end{align}
where (a) follows from switching the order of differentiation and expectation by applying Dominated Convergence Theorem (DCT)~\cite{dct}.
Therefore, the coverage probability conditioned on $R$ and $\ncalU$ can now be written as
\begin{align}\label{equ:pcm3}
\P\big( \sir > \T | R, \ncalU, x_0\big) = \sum_{k=0}^{m_0-1} \frac{(-m_0 \T r^{\alpha})^k}{k!} \bigg[ \frac{\partial^k}{\partial s^k} \bigg[ \prod_{i=1}^{N-1} \bigg(1 + \frac{s U_i^{-\alpha}}{m} \bigg)^{-m} \bigg] \bigg]_{s=m_0 \T r^{\alpha}} .
\end{align}
The results obtained upon applying the limits on the conditional coverage probability are given in the following Theorem.
\begin{thm}
The coverage probability of the receiver for a no-fading channel conditioned on the serving distance $R$ and the set of interfering distances $\ncalU$ is
\begin{equation}
\P\left( \sir > \T | R, \ncalU, x_0\right) = \begin{dcases}
1 \qquad 0<z<1 \\
\frac{1}{2} \qquad z=1 \\
0 \qquad z>1
\end{dcases},
\end{equation}
where $z=\T r^{\alpha} \sum_{i=1}^{N-1}u_i^{-\alpha}$.
\end{thm}
\begin{IEEEproof}
Applying limits $m_0 \to \infty$ and $m \to \infty$ in \eqref{equ:pcm3}, we get the conditional coverage probability as 
\begin{align}\notag
\P\big( \sir > \T | R,\ncalU, x_0\big) &= \lim_{m_0 \to \infty} \lim_{m \to \infty} \sum_{k=0}^{m_0-1} \frac{(-m_0 \T r^{\alpha})^k}{k!} \bigg[ \frac{\partial^k}{\partial s^k} \bigg[ \prod_{i=1}^{N-1} \bigg(1 + \frac{s u_i^{-\alpha}}{m} \bigg)^{-m} \bigg] \bigg]_{s=m_0 \T r^{\alpha}} \\\notag
&\stackrel{(a)}{=} \lim_{m_0 \to \infty}  \sum_{k=0}^{m_0-1} \frac{(-m_0 \T r^{\alpha})^k}{k!} \bigg[ \frac{\partial^k}{\partial s^k} \bigg[ \prod_{i=1}^{N-1} \lim_{m \to \infty} \bigg(1 + \frac{s u_i^{-\alpha}}{m} \bigg)^{-m} \bigg] \bigg]_{s=m_0 \T r^{\alpha}} \\\notag
&= \lim_{m_0 \to \infty}  \sum_{k=0}^{m_0-1} \frac{(-m_0 \T r^{\alpha})^k}{k!} \bigg[ \frac{\partial^k}{\partial s^k} \bigg[ \exp\bigg(-s\sum_{i=1}^{N-1}u_i^{-\alpha}\bigg) \bigg] \bigg]_{s=m_0 \T r^{\alpha}} \\ \label{equ:tayl1}
&\stackrel{(b)}{=} \lim_{m_0 \to \infty}  \sum_{k=0}^{m_0-1} \frac{\big(m_0 z \big)^k}{k!} \bigg[  \exp\big(-m_0 z\big) \bigg] \\\notag
&= \lim_{m_0 \to \infty} \frac{\Gamma(m_0, m_0 z)}{\Gamma(m_0)} ,
\end{align}
where (a) follows from applying the limit before differentiation and writing the limit of product as product of limits, and (b) follows from the $k^{th}$ derivative of exponential function followed by the substitution $z = \beta r^{\alpha} \sum_{i=1}^{N-1} u_i^{-\alpha}$.

Now, we apply the limit $m_0 \to \infty$ to the above equation. Although the expression in \eqref{equ:tayl1} resembles the Taylor series expansion of exponential function, it is not possible to directly take the limit as both the limit of the summation and the summand approach $\infty$. This problem has been well studied in Mathematics~\cite{abe,tricomi,temme,walter} in relation to incomplete gamma function and this sum is asymptotically equal to the following function.
\begin{align}
\frac{\Gamma(m_0, m_0 z)}{\Gamma(m_0)} = \frac{1}{2} - \frac{1}{\sqrt{\pi}} \erf\bigg(\sqrt{\frac{m_0}{2}}(z-1)\bigg) + \frac{1}{3}\sqrt{\frac{2}{m_0 \pi}}\bigg(1+\frac{m_0(z-1)^2}{2}\bigg) e^{-\frac{m_0(z-1)^2}{2}} + O\Big(\frac{1}{m_0}\Big).
\end{align}
For completeness, the proof of this asymptotic expansion is provided in Appendix \ref{app:igf}. 
Applying the limit $m_0 \to \infty$, we see that the limit converges to three values depending on the range of $z$. This completes the proof.
\end{IEEEproof}

We notice that the condition under which the coverage probability converges to 1 is $z<1$, i.e., $\T r^{\alpha} \sum_{i=1}^{N-1}u_i^{-\alpha} < 1$. This is nothing but the condition of $\sir>\T$ in no-fading channels. Therefore, quite interestingly, we obtain only a redundant condition when the coverage probability in no-fading channels is evaluated from the results of Nakagami-$m$ fading. To the best of our understanding, this insight has not been reported before in the context of the limiting case of Nakagami-$m$ fading. In the next subsection, we provide an accurate approximation to compute  coverage probability in the absence of fading.

\subsection{Dominant Interferer Approach}
In this method, we capture the effect of dominant interferer exactly and approximate the aggregate interference from rest of the interferers to a Gaussian random variable.
The $\sir$ at the receiver in the absence of fading is
\begin{equation}
\sir = \frac{R^{-\alpha}}{\sum_{i=1}^{N-1} U_i^{-\alpha}} = \frac{R^{-\alpha}}{U_1^{-\alpha} + \sum_{i=2}^{N-1} U_i^{-\alpha}}.
\end{equation}
Let $I_{N-2} = \sum_{i=2}^{N-1} U_i^{-\alpha}$. Since the distance of the interferers from the receiver $U_i$ conditioned on $R$ and $U_1$ are i.i.d., the terms $U_i^{-\alpha}$ that constitute the sum $I_{N-2}$ are also conditionally i.i.d. Therefore, by central limit theorem (CLT), the sum of i.i.d. random variables $I_{N-2}$ can be approximated by a normal random variable, whose mean and variance are given by the following Lemmas.

\begin{lemma}\label{lem:eI}
The conditional mean of the interference power at the receiver excluding the interference from the dominant interferer is 
\begin{align}
\nbbE\big[I_{N-2}| R, U_1, x_0 \big] = 
\begin{dcases}
\frac{2(N-2) \big({w_m}^{2-\alpha} - u_1^{2-\alpha} \big)}{(2-\alpha)(d^2- u_1^2)} + \\
\qquad (N-2)\int_{w_m}^{w_p} u_i^{-\alpha}\frac{f_{W_{i,2}}(u_i|x_0)}{1-F_{W_{i,1}}(u_1|x_0)} {\rm d}u_i, &h \leq u_1 \leq w_m \\
(N-2)\int_{w_m}^{w_p} u_i^{-\alpha}\frac{f_{W_{i,2}}(u_i|x_0)}{1-F_{W_{i,2}}(u_1|x_0)} {\rm d}u_i,      &w_m \leq u_1 \leq w_p
\end{dcases}.
\end{align}
\end{lemma}
\begin{IEEEproof}
The mean interference (excluding interference from the dominant interferer) conditioned on the serving distance $R$ and the distance to the dominant interferer $U_1$ is given by
\begin{align}
\nbbE\big[ I_{N-2} | R, U_1, x_0 \big] = (N-2) \nbbE \big[ U_i | R, U_1, x_0 \big ].
\end{align}
This follows from the conditionally i.i.d. nature of the distances $U_i$. Using the conditional distribution of $U_i$ derived in Lemma \ref{lem:pdfui} and solving the resulting integral gives the final result. This completes the proof.
\end{IEEEproof}

\begin{lemma}\label{lem:varI}
The conditional variance of interference power at the receiver excluding the interference from the dominant interferer is 
\begin{align}\notag
\V\big[I_{N-2}| R, U_1, x_0 \big] &= (N-2)\Bigg[\int_h^{w_p}u_i^{-2\alpha}f_{U_i}(u_i | r, u_1, x_0)  {\rm d}u_i \\
&\qquad \qquad-  \bigg(\int_h^{w_p}u_i^{-\alpha}f_{U_i}(u_i | r, u_1, x_0)  {\rm d}u_i \bigg)^2 \Bigg].
\end{align}
\end{lemma}
\begin{IEEEproof}
The proof follows from the definition of variance and conditionally i.i.d. distances $U_i$ whose distribution is given in Lemma \ref{lem:pdfui}. 
\end{IEEEproof}
While these integrals can not be reduced to closed-form, it is easy to evaluate them numerically. However, for a receiver located at the origin, these expressions can be simplified to closed-form expressions given in the following Corollaries.
\begin{cor}
The conditional mean of the interference power experienced by the receiver at the origin excluding the interference from dominant interferer is 
\begin{align}
\nbbE[I_{N-2}|R, U_1] = \frac{2(N-2)[u_1^{2-\alpha}-d^{2-\alpha}]}{(\alpha-2)(d^2-u_1^2)}. 
\end{align}
\end{cor}
\begin{cor}
The conditional variance of the interference power experienced by the receiver located at the origin excluding the interference from dominant interferer is 
\begin{align}
 \V[I_{N-2} | R, U_1] = (N-2)\Bigg[- \frac{4(d^{2 - \alpha} - u_1^{2 - \alpha})^2}{(d^2 - u_1^2)^2(\alpha - 2)^2} - \frac{(d^{2 - 2\alpha} - u_1^{2 - 2\alpha})}{(d^2 - u_1^2)(\alpha - 1)}\Bigg].
\end{align}
\end{cor}
Using these results, we derive an accurate approximation for coverage probability in the following Theorem.
\begin{thm}\label{thm:clt} 
The coverage probability of the receiver can be approximated using dominant-interferer approach as
\begin{align}\notag
P_c \approx \bigintsss_h^{w_p} \bigintsss_{r}^{w_p} \Bigg[ 1 - Q\Bigg( \frac{\beta^{-1} r^{-\alpha} - u_1^{-\alpha} - \mu_{I_{N-2}}}{\sigma_{I_{N-2}}} \Bigg) \Bigg] f_{R,U_1}(r,u_1 | x_0) {\rm d}u_1 {\rm d}r,
\end{align}
where $\mu_{I_{N-2}}$ and $\sigma_{I_{N-2}}^2$ are mean and variance of interference given by Lemmas \ref{lem:eI} and \ref{lem:varI}, respectively. Q($\cdot$) is the Q-function.
\end{thm}
\begin{IEEEproof}
The coverage probability is given by
\begin{equation}\label{eq:pc_gen2}
P_c = \int_h^{w_p} \int_{r}^{w_p} \P(\sir> \beta | R,U_1,x_0)f_{R,U_1}(r,u_1| x_0) {\rm d}u_1 {\rm d}r,
\end{equation}
where the probability term in the integrand is
\begin{equation}
\P(\sir > \beta | R, U_1) = \P\Big(I_{N-2} < \beta^{-1} R^{-\alpha} - U_1^{-\alpha} \Big).
\end{equation}
As stated earlier, $I_{N-2}$ is the sum of i.i.d. random variables. Therefore, by applying CLT, the above probability is given by the CDF of a Gaussian random variable:
\begin{equation}
\P(\sir> \beta | R, U_1) = 1 - Q \Bigg( \frac{\beta^{-1} r^{-\alpha} - u_1^{-\alpha} - \mu_{I_{N-2}}}{\sigma_{I_{N-2}}} \Bigg),
\end{equation}
where $\mu_{I_{N-2}}$ and $\sigma_{I_{N-2}}^2$ are the mean and variance of $I_{N-2}$, as given in Lemmas \ref{lem:eI} and \ref{lem:varI}, respectively. Substituting the above result and the joint distance distribution from Lemma \ref{lem:jpdfu1u2} in \eqref{eq:pc_gen2}, we obtain the coverage probability.
\end{IEEEproof}

\subsection{Bounds of Coverage Probability Approximation}
In this subsection, using BET, we analyze the tightness of the coverage probability approximation proposed in the previous subsection. BET gives a bound on the maximal deviation of the normal distribution from the true distribution in terms of the moments of the distribution. By BET, for a sequence of random variables $X_1, X_2,....X_n$, with $\nbbE[X_i | R, U_1, x_0] = 0,\  \nbbE[X_i^2 |R, U_1, x_0]= \sigma^2$,  $\nbbE[ |X_i|^3 | R, U_1, x_0] = \rho$, and sample mean $M_n = \frac{1}{n}\sum_{i=1}^n X_i$, the error between the actual distribution $F_n(x)$ of the random variable $\frac{M_n\sqrt{n}}{\sigma}$ and the standard normal distribution $\Phi_n(x)$ is bounded by $\frac{C \rho}{\sigma^3 \sqrt{n}}$, i.e.,
\begin{equation}
|F_n(x) - \Phi_n(x)| \leq \frac{C \rho}{\sigma^3 \sqrt{n}}
\end{equation}
where $C$ is a constant. The best known estimate of $C$ is $C<0.4748$~\cite{shevtsova}. We now rewrite the expression for interference in a form that will allow us to conveniently apply BET. Let $V_i = U_i^{-\alpha}$ and $X_i = V_i - \mu_{V_i}$ where $\mu_{V_i}$ is the mean interference from Lemma \ref{lem:eI}. Therefore, we have $\nbbE[X_i | R, U_1]=0$ and $\nbbE[X_i^2 | R, U_1, x_0] = \V[U_i^{-\alpha}|R, U_1, x_0]$, which is given by Lemma \ref{lem:varI}. The third moment of absolute value of $X_i$, $\rho$, is computed in the following Lemma.
\begin{lemma}\label{lem:rho}
The third moment of absolute value of $X_i$ conditioned on $R$ and $U_1$ is
\begin{align}
\nbbE \big[ |X_i|^3 | U_1, x_0 \big] = \begin{dcases}
\int_{w_p^{-\alpha}-\mu_{V_i}}^0 -x_i^3 f_{U_i}\big( (x_i + \mu_{V_i})^{-1/\alpha} | r, u_1, x_0\big) {\rm d}x_i \\
\quad + \int_0^{u_1^{-\alpha} - \mu_{V_i}} x_i^3 f_{U_i}\big( (x_i + \mu_{V_i})^{-1/\alpha} |r,u_1,x_0\big) {\rm d}x_i, \quad &u_1^{-\alpha} - \mu_{V_i} \geq 0 \\
\int_{w_p^{-\alpha}-\mu_{V_i}}^{u_1^{-\alpha} - \mu_{V_i}} -x_i^3 f_{U_i}\big( (x_i + \mu_{V_i})^{-1/\alpha} |r,u_1,x_0\big){\rm d}x_i, &u_1^{-\alpha} - \mu_{V_i} < 0 
\end{dcases}.
\end{align}
\end{lemma}
\begin{IEEEproof}
See Appendix \ref{app:rho}.
\end{IEEEproof}
While it is difficult to get a simple closed-form expression for the above result, it can be easily evaluated numerically. Using this result, along with the first and second moments of $X_i$, we derive the bounds of coverage probability in the following Theorem.

\begin{thm}\label{thm:bounds}
The coverage probabilty $P_c$ of the receiver, at a distance $x_0$ from the origin, is bounded as
\begin{align*}
P_l \leq \pc \leq P_u
\end{align*}
with
\begin{align}
P_l = \bigintsss_h^{w_p} \bigintsss_r^{w_p} \bigg[\Phi \Big( \calG\big(\T, \alpha, N, R,U_1, x_0\big)  \Big) - \frac{C\rho}{\sigma^3 \sqrt{N-2}} \bigg] f_{R, U_1}(r, u_1|x_0) {\rm d}u_1 {\rm d}r, \\
P_u = \bigintsss_h^{w_p} \bigintsss_r^{w_p} \bigg[\Phi \Big( \calG\big(\T, \alpha, N, R,U_1, x_0\big) \Big) + \frac{C\rho}{\sigma^3 \sqrt{N-2}} \bigg] f_{R, U_1}(r, u_1|x_0) {\rm d}u_1 {\rm d}r,
\end{align}
where $\Phi(\cdot)$ is the CDF of standard normal distribution, $C = 0.4748$, $\sigma^2$ is the variance given in Lemma \ref{lem:varI} and $\rho$ is the third moment given by Lemma \ref{lem:rho}.
\end{thm}
\begin{IEEEproof}
See Appendix \ref{app:bounds}.
\end{IEEEproof}
While these bounds are loose for very small values of $N$, we observe that the error between the normal distribution and true distribution decreases at the rate $(N-2)^{-\frac{1}{2}}$. In other words, as the number of transmitting devices grows to a large value, the approximate coverage probability converges to the actual value. However, in the next section, we show that this approximation is surprisingly accurate even for a small number of nodes in the network.

\section{Results and Discussion}
In this section, we validate our analytical results by comparing the theoretical coverage probabilities with the simulation results for a finite network of UAVs. In addition to providing useful design insights, we also discuss the applicability of our proposed analytic approaches to an urban setting in which the visibility of some UAVs is obstructed/blocked by buildings.

\subsection{Numerical Results}
\begin{figure}
\begin{minipage}{.49\textwidth}
\includegraphics[width=1\textwidth]{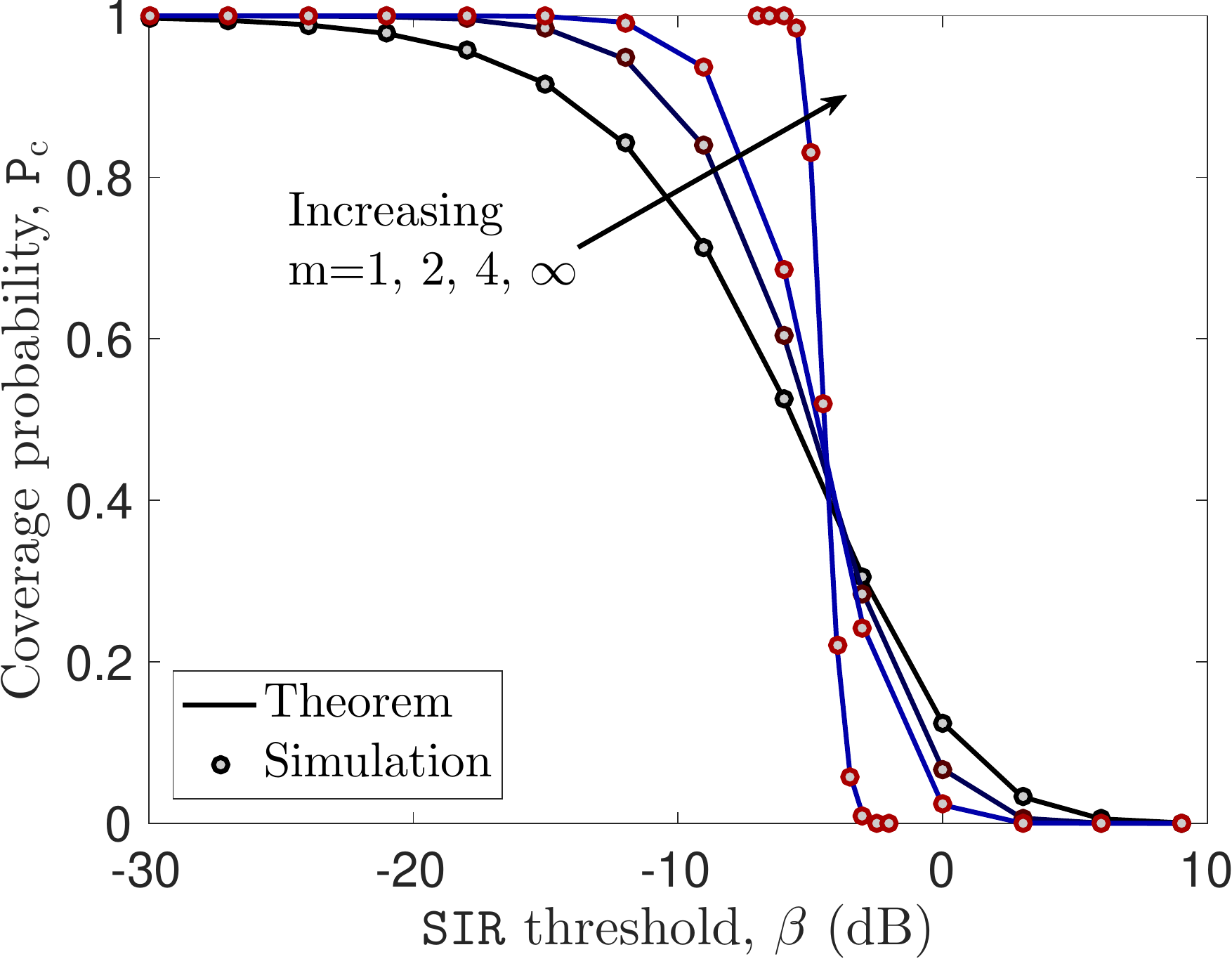}
\caption{Coverage probability of the receiver as a function of $\sir$ threshold ($h=10$ km, $x_0 = 4$ km, and $\alpha = 2.5$).}
\label{fig:impactm}
\end{minipage}%
\hfill
\begin{minipage}{.49\textwidth}
\includegraphics[width=1\textwidth]{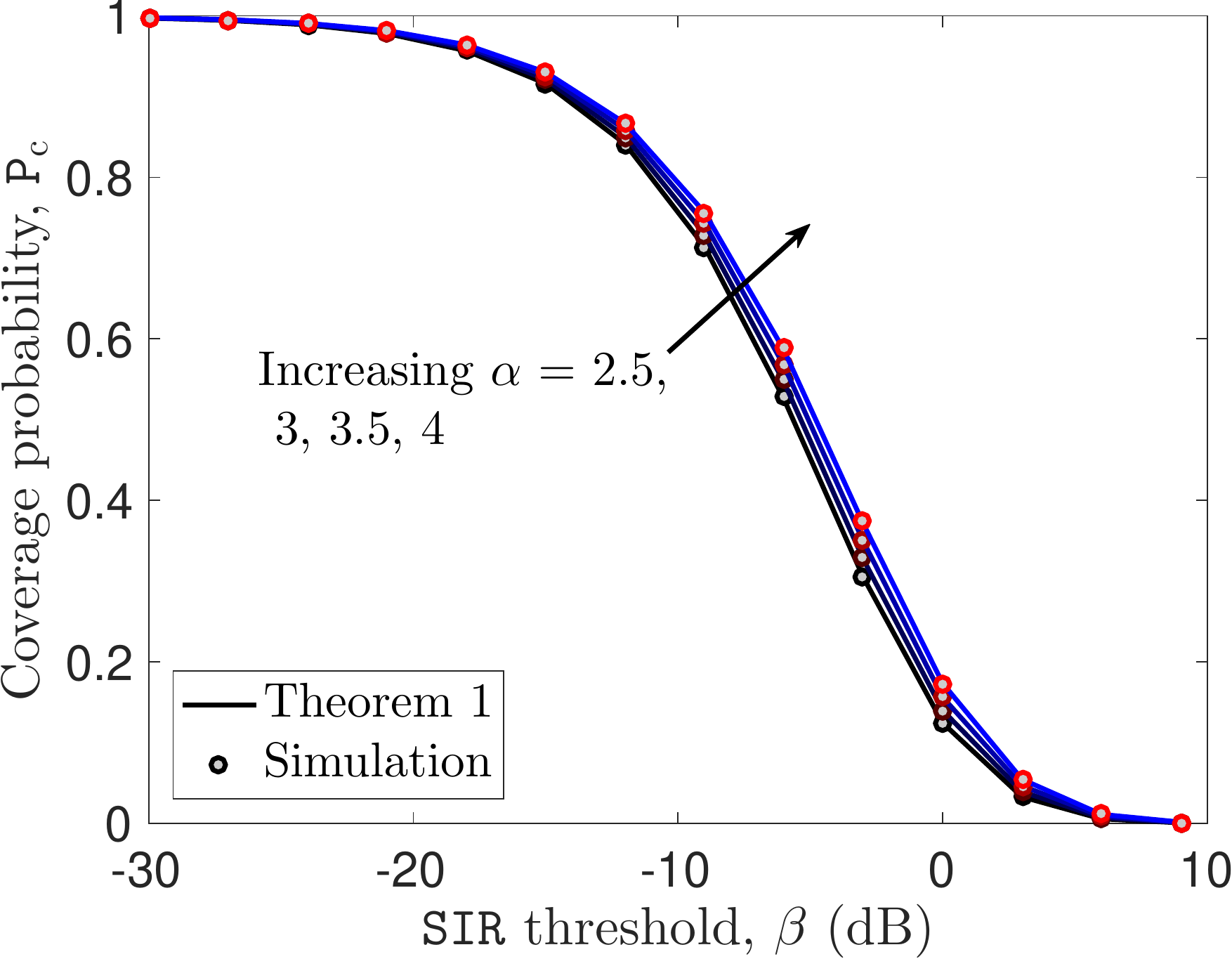}
\caption{Coverage probability of the receiver as a function of $\sir$ threshold ($h = 10$ km, $m = 1$, and $x_0 = 4$ km).}
\label{fig:impacta}
\end{minipage}
\end{figure}

We simulate a finite network of UAVs with $N=5$, uniformly distributed in a disk of radius $r_a = 10$ km. We evaluate the coverage probability for different system parameters and compare them with the theoretical results obtained in Theorems \ref{thm:pcm} and \ref{thm:clt}. We find that our theoretical results match exactly with the simulations as shown in Fig. \ref{fig:impactm}. The key factors that affect the coverage probability are: (i) channel fading parameter $m$, (ii) path-loss exponent $\alpha$, (iii) the height of UAVs $h$,  and (iv) the distance of the receiver from the origin $x_0$. We study the impact of each parameter on the coverage probability in the rest of this Subsection.

{\em Impact of fading}. We compute the coverage probability of the receiver as a function of SIR threshold $\beta$ for $m = 1,\ 2,\ 4$ and $\infty$. Note that $m \to \infty$ is nothing but the no-fading scenario. The other parameters of the simulation are $h=10$ km, $x_0 = 4$ km, and $\alpha = 2.5$. As expected, the variance of SIR decreases with the increase in $m$. In other words, SIR starts {\em concentrating} as we move from Rayleigh fading ($m=1$) case to the no-fading ($m\rightarrow \infty$) case. 

{\em Impact of path-loss exponent}. We study the impact of path-loss exponent $\alpha$ on coverage in Fig. \ref{fig:impacta}, where we plot the coverage probability as a function of $\sir$ threshold $\T$ for different values of $\alpha$. The simulations are run for $h = 10$ km, $m = 1$, and $x_0 = 4$ km. It can be observed that the coverage probability degrades with a decrease in the path-loss exponent. While reducing $\alpha$ increases the received power of the desired signal, it also increases the interference power, thereby degrading the overall SIR and hence the coverage probability.

\begin{figure}
\begin{minipage}{.49\textwidth}
\includegraphics[width=1\textwidth]{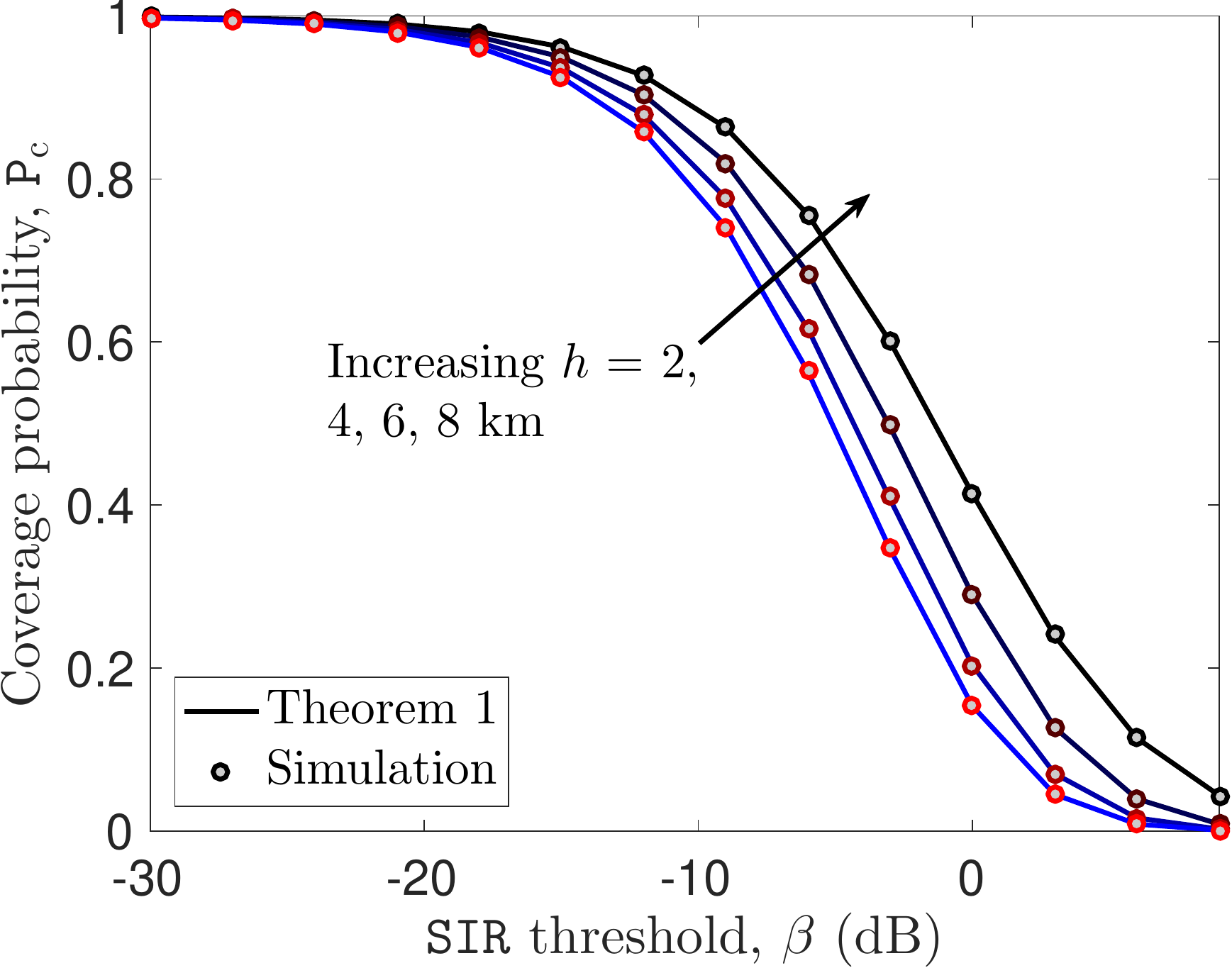}
\caption{Coverage probability of the receiver as a function of $\sir$ threshold ($m=1$, $\alpha = 2.5$, and $x_0= 1$ km).}
\label{fig:impacth}
\end{minipage}%
\hfill
\begin{minipage}{.49\textwidth}
\includegraphics[width=1\textwidth]{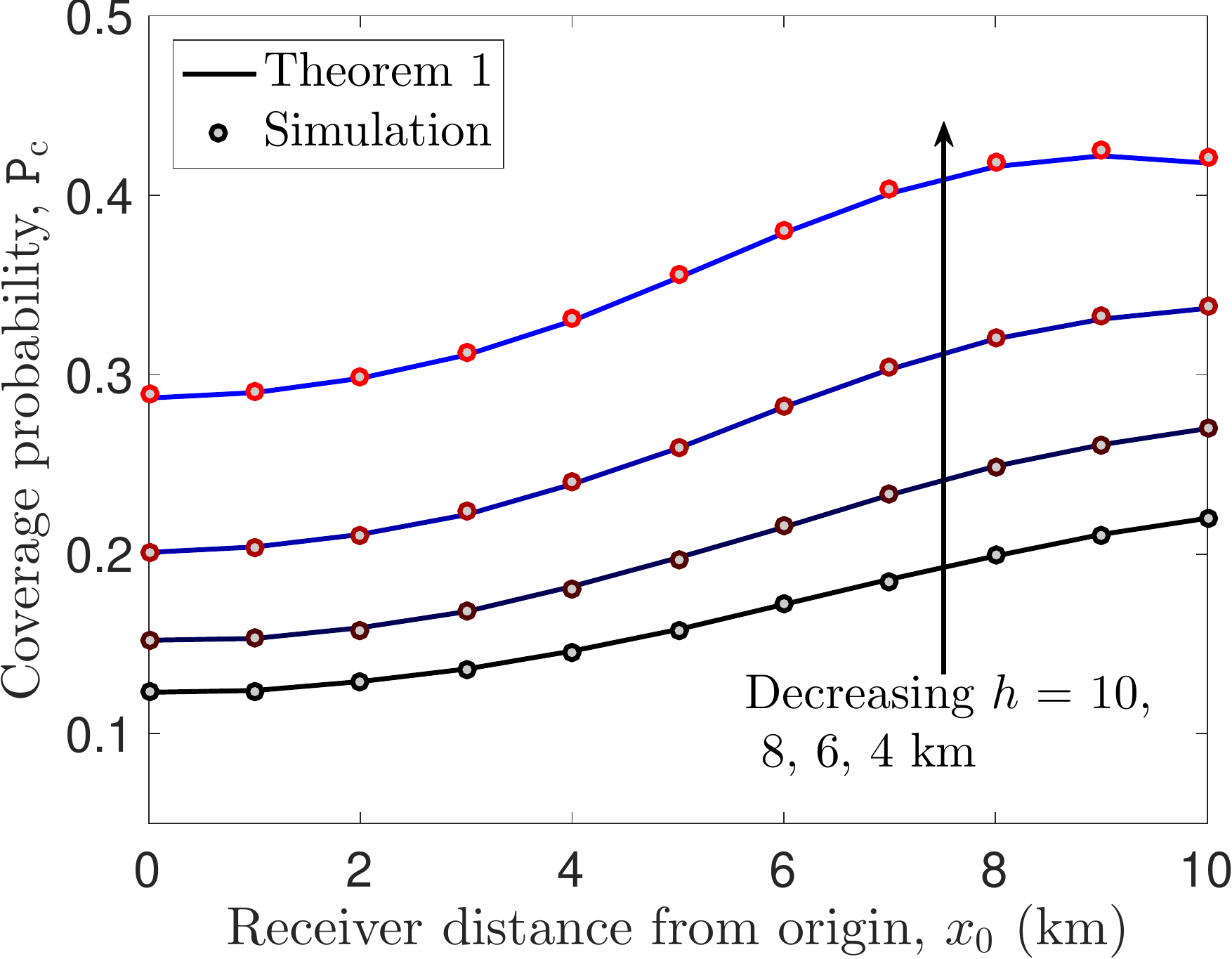}
\caption{Coverage probability as a function of receiver distance from the origin ($m=1$, $r_a = 10$ km, $\alpha=2.5$, and $N =5$).}
\label{fig:impactx0}
\end{minipage}
\end{figure}

{\em Impact of height}. We compare the coverage probability of the receiver as a function of $\sir$ threshold $\beta$ for different values of the height of UAVs (2, 4, 6, and 8 km) . The other simulation parameters were fixed at $m=1$, $\alpha = 2.5$, and $x_0= 1$ km. It can be observed from Fig. \ref{fig:impacth} that the coverage probability deteriorates as the height $h$ of the UAVs increases. An increase in the height of UAVs increases the distance between the receiver and the transmitters. Intuitively, when viewed from a receiver that moves away from the transmitters, the separation between the serving and interfering nodes tends to diminish. This worsens the SIR and hence the coverage.

{\em Impact of receiver distance from the origin}. The impact of receiver distance from the origin $x_0$ on coverage probability can be studied from Fig.  \ref{fig:impactx0} where we plot coverage as a function of $x_0$ for different values of $h$. The other simulation parameters were $r_a = 10$ km, $\alpha=2.5$, $\beta=0$ dB, and $N =5$. It can be observed that the coverage probability varies significantly with the location of the receiver, which highlights the importance of assuming arbitrarily located receiver.%
\subsection{Applicability of the Proposed Results to a Relevant Urban Model}
\begin{figure}
\begin{minipage}{.48\textwidth}
\includegraphics[width=1\textwidth]{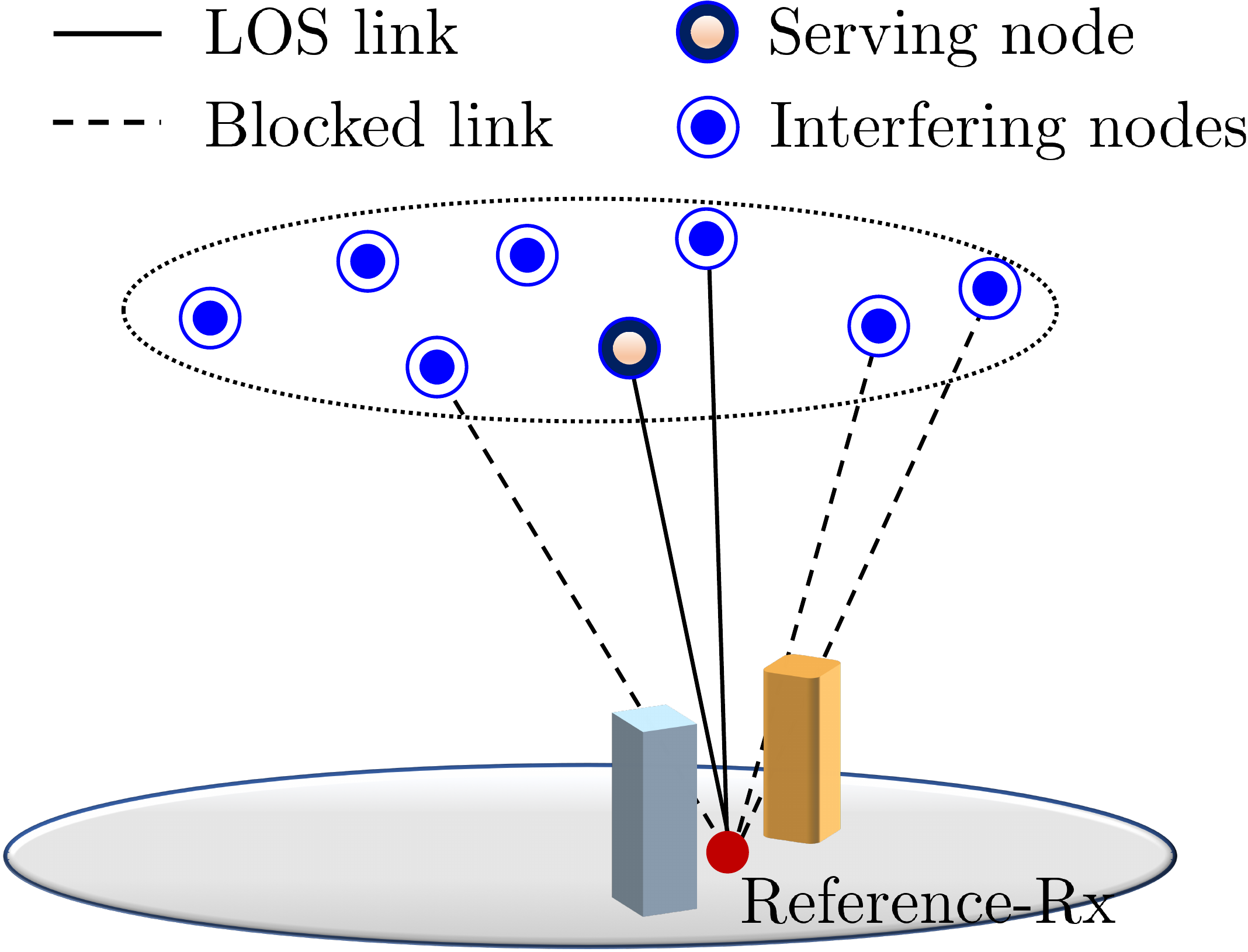}
\caption{Illustration of blockage model}
\label{fig:sysmod4}
\end{minipage}%
\hfill
\begin{minipage}{.48\textwidth}
\includegraphics[width=1\textwidth]{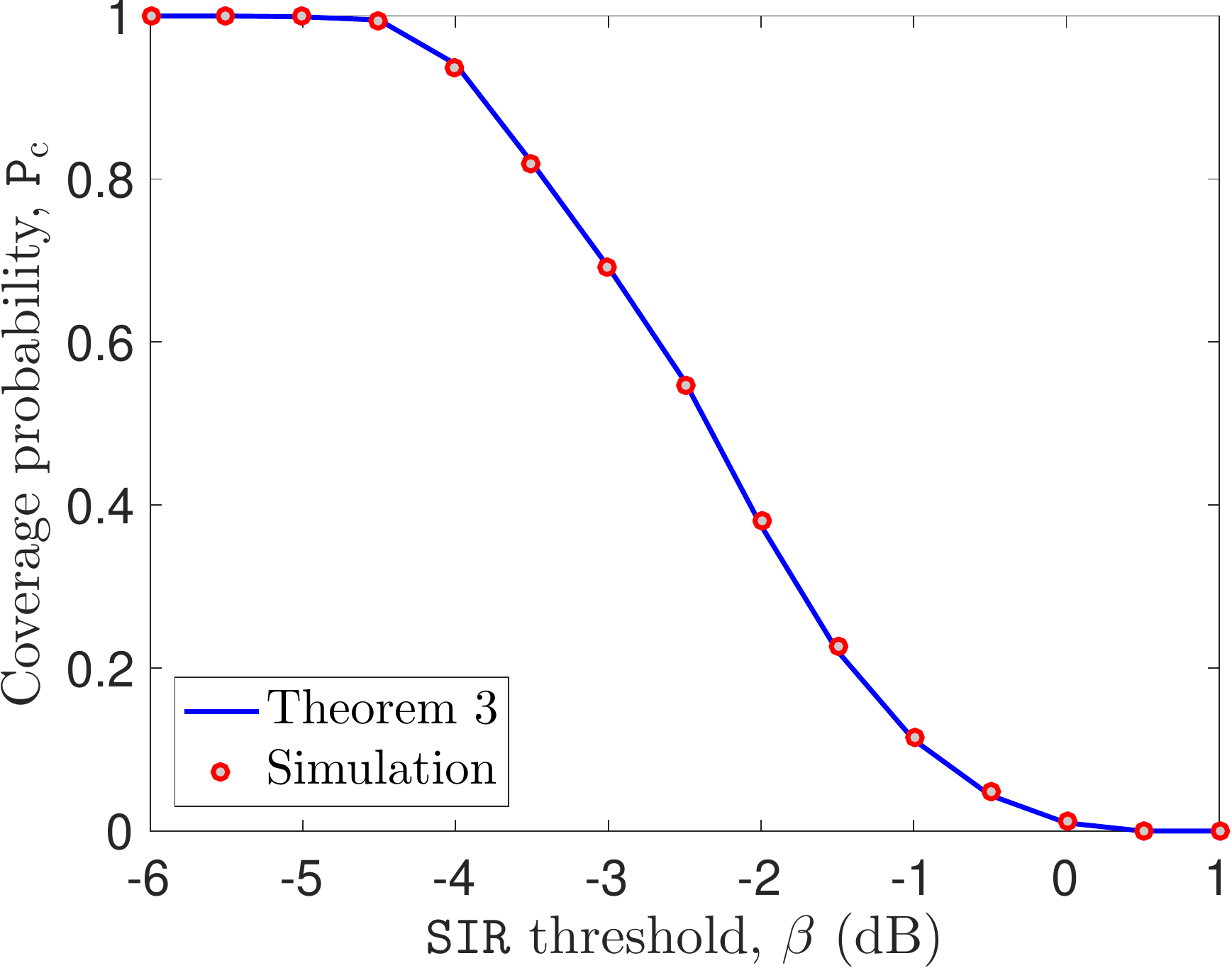}
\caption{Coverage Probability as a function of SIR threshold ($h=10$ km, $x_0 = 1$ km and $\alpha = 2.5$ ).}
\label{fig:impactb}
\end{minipage}
\end{figure}

As discussed already, the main technical objective of this paper was to develop a comprehensive framework for the analysis of a reasonable canonical model for finite UAV networks. Before concluding this Section, we demonstrate that this canonical model can be enriched to some extent while retaining its tractability. In particular, we incorporate the effect of shadowing/blocking due to high-rises that will be prominent in urban regions, especially at higher frequencies. The system setup is illustrated in Fig. \ref{fig:sysmod4}. The blockages may result in attenuation of the received signal and hence affect the coverage probability. We begin with the spatial model considered in this paper, where $N$ transmitting devices are uniformly distributed in a disk $b(o',r_a)$ at an altitude $h$ above the ground. Due to the presence of buildings, all the UAVs may not be {\em visible} at the receiver. We denote the number of visible UAVs by $N_v \leq N$. In our {\em analytical} treatment, we assume that the LOS paths from the receiver to each UAV are blocked independently of each other. In other words, we ignore the correlation in blocking introduced by the spatial distribution of buildings in the area. As a result, $N_v$ can be modeled as a binomial random variable. This {\em independence} assumption will be validated through a numerical comparison in Fig.~\ref{fig:impactb}. 

For simplicity, we limit our discussion to direct path propagation and ignore multi-path fading. We assume that if a UAV is hidden behind a building, its signal is attenuated by a fixed factor $\eta$. While we can easily extend this discussion to a more general setup, this simple scenario is sufficient to fix the key ideas. For this setup, the received signal power $P_i$ from the $i^{th}$ transmitter, located at a distance $W_i$ from the receiver is
$P_i = B_i W_i^{-\alpha}$, 
where $B_i = 1$ if the UAV is visible at the receiver and $\eta < 1$ otherwise. Recall that the distribution of $W_i$ is given by Lemma \ref{lem:pdfwi}. Now conditioned on $N_v = n_v$, we get two independent BPPs: (i) a BPP formed by $n_v$ visible UAVs, and (ii) a BPP formed by $N-n_v$ blocked UAVs. Conditional on $N_v = n_v$, the coverage analysis can be performed following the proposed approach. We do not go into the mathematical details due to lack of space. Note that if $\eta=0$, we get only one BPP (of visible UAVs), which reduces this setup to that of Theorem~\ref{thm:clt}, which will be used for numerical comparisons below.


The key approximation made in the above analysis is the {\em independent blocking} assumption. We validate this assumption numerically. In particular, we simulate a urban scenario in MATLAB with 5 UAVs uniformly distributed over a circular area of radius of 10 km. We assume a uniform distribution of 50 buildings that are of dimensions $50$m $\times\ 50$m $\times\ 150$m. We choose the simple case of $\eta=0$, i.e., we receive signals from only those UAVs that are visible. For a receiver at a distance of 1 km from the origin and path-loss exponent of $\alpha = 2.5$, we obtain the coverage probability from simulations. For each realization, we numerically obtain $n_v$, which is used instead of $N$ in Theorem~\ref{thm:clt} to obtain the conditional coverage probability analytically. Monte-Carlo simulations are used to average over $N_v$. Note that the purpose of this comparison is to show that the independent blocking assumption is reasonable. This is quite evident in the results presented in Fig. \ref{fig:impactb}, where the simulation results are the ones obtained from actual numerical experiments without any assumptions, and the analytical result is obtained under independent blocking assumption. {\em This discussion shows that the canonical setup introduced in this paper can be extended in many meaningful ways to study various aspects of UAV networks.}

\section{Conclusion}
In this paper, we have presented a comprehensive downlink coverage analysis for a finite three-dimensional wireless network formed by $N$ UAVs. Modeling the network of UAVs as a BPP, we characterized the distribution of distances from the receiver to the serving and interfering nodes. We first derived an exact expression for coverage probability for the reference receiver under independent Nakagami-$m$ fading channels in terms of the derivatives of the Laplace transform of interference power distribution. Using asymptotic expansion of incomplete gamma function, we showed that the coverage probability for an important special case of no-fading cannot be obtained explicitly as the limiting case of Nakagami-$m$ fading by taking the limit $m \to \infty$. As a result, we developed an alternate approach to compute the approximate coverage probability in which the interference from dominant interferer is modeled exactly and the distribution of residual interference from other interferers is approximated to a normal distribution by CLT. We then obtained the bounds of the approximate coverage probability using  Berry-Esseen theorem, which quantifies the rate of convergence of the normal approximation to the true distribution. Our analysis revealed several useful performance trends in terms of the heights of the UAVs and channel propagation characteristics. We also discussed a possible extension of the proposed canonical model to a simple urban scenario in which the the UAVs are shadowed by high-rises. 

This work has numerous extensions. The mathematical tools developed in the paper can be applied to the analysis of more general three-dimensional finite networks. The setup studied in this paper can also be extended to study the co-existence of UAV networks and terrestrial cellular networks. From modeling perspective, a useful direction of work wold be to develop more sophisticated but tractable three-dimensional spatial models that account for the curvature of the earth. In terms of performance evaluation, the proposed framework can be used to study other useful metrics besides coverage, such as throughput and energy efficiency.


\appendix
\subsection{Proof of Lemma \ref{lem:cdfwi}}\label{app:cdfwi}
The cumulative distribution function (CDF) of each element of the sequence $\{W_i\}$ is 
\begin{align}\notag  
F_{W_i}(w_i) &= \P(W_i \leq w_i) = \P(S_i^2 + h^2 \leq w_i^2) \\ 
&= F_{S_i}\Big(\sqrt{w_i^2-h^2} \Big), \label{eq:lemfw}
\end{align}
where $F_{S_i}(s_i)$ is the CDF of the distance between the receiver and the projection of the location of $i^{th}$ transmitter onto the ground plane. The projections of the locations of the transmitters form a two dimensional BPP on the ground plane. Using the approach presented in~\cite{mathai1999introduction}, the conditional CDF of $S_i$ is computed as the the area of intersection of the disks $b(o,r_a)$ and $b({\bf{x}},s_i)$, divided by the area of the disk $b(o,r_a)$. Depending on the range of $s_i$, there are two possible cases: (i) the disk $b({\bf{x}},s_i)$ is entirely contained in the disk $b(o,r_a)$, and (ii) $b({\bf{x}},s_i)$ partially overlaps with $b(o,r_a)$. Therefore, we obtain a piece-wise conditional CDF of $S_i$ as given below: 
\begin{align}\label{eq:cdfbpp}
F_{S_i}(s_i|x_0) = 
\begin{dcases}
	 F_{S_{i,1}}(s_i | x_0 ), &    0 \leq s_i \leq s_m \\
	 F_{S_{i,2}}(s_i | x_0 ), &    s_m < s_i \leq s_p
\end{dcases},
\end{align}
with
\begin{align}
&F_{S_{i,1}}(s_i | x_0 ) =\frac{s_i^2}{r_a^2}, \quad F_{S_{i,2}}(s_i | x_0 ) =\frac{s_i^2}{\pi r_a^2}(\theta^* - \frac{1}{2}\sin 2\theta^*) + \frac{1}{\pi}(\phi^* - \frac{1}{2}\sin 2\phi^*),  
\intertext{where} \ \notag &\theta^* = \arccos\bigg(\frac{s_i^2+x_0^2-r_a^2}{2 x_0 s_i }\bigg), \phi^* = \arccos\bigg(\frac{x_0^2+r_a^2-s_i^2}{2 x_0 r_a}\bigg), s_m = r_a-x_0, \ \text{and} \ s_p = r_a+x_0. 
\end{align}

Substituting \eqref{eq:cdfbpp} in \eqref{eq:lemfw}, we obtain the conditional CDF of $W_i$. This completes the proof. 

\subsection{Proof of Lemma \ref{lem:pdfu1}}\label{app:pdfu1}
Since the receiver connects to the closest transmitter, the serving distance $R$ is by definition $R= \min\{W_i\} $, where the distribution of $W_i$ is given in Lemma \ref{lem:cdfwi}. The conditional CDF of $R$ can therefore be computed as 
\begin{align}\notag
F_{R}(r|x_0) = \P(R \leq r | x_0) = 1 - \P \big( \min\{W_i\} > r|x_0 \big) &= 1-\P \big(W_1 > r, W_2 > r, ... , W_{N} > r |x_0 \big) \\ 
&\stackrel{(a)}{=} 1-\big( 1 - F_{W_i}(r|x_0) \big)^{N},
\end{align}
where (a) follows from the i.i.d nature of the set of distances ${W_i}$.
Differentiating the above expression w.r.t. $r$, the PDF of the serving distance is obtained as
\begin{align}
f_{R}(r|x_0) =  N\big( 1 - F_{W_i}(r|x_0) \big)^{N-1}f_{W_i}(r|x_0).
\end{align}
The PDF of $R$ can be obtained by substituting the results from Lemmas \ref{lem:cdfwi} and \ref{lem:pdfwi} in the above equation. This completes the proof.
\subsection{Proof of Lemma \ref{lem:pdfuigu1}}\label{app:pdfuigu1}
The joint density function of the ordered subset $\{W_{(i)}\}_{i=2:N}$ conditioned on the serving distance $R$, and $x_0$ is 
\begin{align}\notag
f(w_{(2)}, w_{(3)}, ... , w_{(N)} | r,  x_0) \stackrel{(a)}{=} \frac{N! f_{W_{i}}(r|x_0) \prod_{i=2}^{N}f_{W_i}(w_i|x_0)}{f_{R}(r | x_0)} \stackrel{(b)}{=} (N-1)! \prod_{i=2}^{N}\frac{f_{W_i}(w_i|x_0)}{1-F_{W_{i}}(r | x_0)} ,
\end{align}
where (a) follows from the joint density function for the order statistics of a sample of size $N$ drawn from the distribution of $W_i$, and (b) follows from the result derived in Lemma \ref{lem:pdfu1}. Following the same argument presented in Lemma 3 in~\cite{mehrnazbpp}, we can say that $(N-1)!$ indicates all possible permutations of the elements in the ordered set $\{W_{(i)}\}_{i=2:N}$. Hence, by the joint density function for ordered set, the unordered set of distances are i.i.d. with PDF $\frac{f_{W_i}(w_i|x_0)}{1-F_{W_{i}}(r | x_0)}$.
\subsection{Proof of Lemma \ref{lem:jpdfu1u2}}\label{app:jpdfu1u2}
For a sequence of i.i.d. random variables, $\{X_i\}_{i=1:n}$, with each element characterized by PDF $f_X(x)$ and CDF $F_X(x)$, the order statistics $\{X_{(i)}\}_{i=1:n}$ are random variables defined by sorting the realizations of the sequence in the increasing order. By order statistics~\cite{ahsanullah2005order}, the joint PDF of the smallest two random variables is 
\begin{align}\label{eq:ord}
f_{X_{(1)},X_{(2)}}(x_1,x_2) = n(n-1)[1-F_X(x_2)]^{n-2}f_X(x_1)f_X(x_2).
\end{align}
In our case, the serving distance $R$ and the distance of the dominant interferer from the receiver $U_1$ are the smallest distances in the set $\{W_i\}$. The joint PDF of the two distances is obtained by substituting the results from Lemmas \ref{lem:cdfwi} and \ref{lem:pdfwi} in \eqref{eq:ord}. This completes the proof. 
\subsection{Proof of Lemma \ref{lem:LI}}\label{app:LI}
The Laplace transform of interference power distribution conditioned on serving distance $R$ can be derived as follows:
\begin{align}\notag
&\calL_I\big(s | r, x_0\big) = \nbbE_I \Big[ \exp{(-sI)} \big| R, x_0 \Big] = \nbbE_I\Bigg[ \exp\bigg(-s \sum_{i=1}^{N-1} G_i U_i^{-\alpha} \bigg) \bigg| R, x_0 \Bigg] \\ \notag
&\stackrel{(a)}{=} \nbbE_U \ \nbbE_G \Bigg[ \prod_{i=1}^{N-1} \exp \Big( -s G_i U_i^{-\alpha} \Big)\bigg| R, x_0 \Bigg] \stackrel{(b)}{=} \nbbE_U  \Bigg[ \prod_{i=1}^{N-1} \nbbE_{G_i} \bigg( \exp \Big( -s G_i U_i^{-\alpha}  \Big) \bigg) \bigg| R, x_0 \Bigg] \\ \label{equ:pcm1}
&\stackrel{(c)}{=} \nbbE_U \Bigg[ \prod_{i=1}^{N-1} \bigg(1 + \frac{s U_i^{-\alpha}}{m} \bigg)^{-m} \bigg| R, x_0 \Bigg] \stackrel{(d)}{=} \Bigg[ \nbbE_{U_i}\bigg[ \bigg(1 + \frac{s U_i^{-\alpha}}{m} \bigg)^{-m} \bigg| R, x_0 \bigg] \Bigg]^{N-1},
\end{align}
where (a) follows from the independence of channel gains and the distances of interferers from the receiver, (b) follows from rewriting the expectation of product as the product of expectation owing to i.i.d. channel gains $\{G_i\}$, (c) follows from the moment generating function (MGF) of gamma random variable $G_i$, and (d) follows from conditionally i.i.d. distances of interferers from the receiver. Now, by applying the definition of mean and using the conditional PDF of $U_i$ from Lemma \ref{lem:pdfuigu1}, we get the final result. This completes the proof.

\subsection{Proof of Asymptotic Expansion of Incomplete Gamma Function}\label{app:igf}
In order to prove the desired result, we first asymptotically estimate the following auxiliary integral as $\alpha \to \infty$,
\begin{align}\label{eq:Iab}
I(\alpha, \beta) = \int_0^\beta e^{\alpha t }(1-t)^\alpha {\rm d}t
\end{align}
where $\alpha, \beta$ are taken to be real for simplicity.
Using Taylor series expansion, we have
\begin{align*}
e^{\alpha t }(1-t)^\alpha = \exp\bigg( \alpha(t + \log(1-t)) \bigg) &= \exp \Bigg[ \frac{-\alpha t^2}{2} -\alpha t^3\bigg(\frac{1}{3} + \frac{t}{4} + \frac{t^2}{5} + ...\bigg) \Bigg] \\
&= e^{-\frac{\alpha t^2}{2}} \sum_{m=0}^{\infty} \frac{(-\alpha)^m}{m!}  t^{3m} \bigg(\frac{1}{3} + \frac{t}{4} + \frac{t^2}{5} + ...\bigg)^m .
\end{align*}
From multinomial expansion, it follows that
\begin{align*}
\bigg(\frac{1}{3} + \frac{z}{4} + \frac{z^2}{5} + ...\bigg)^m = \sum_{k=0}^{\infty} a_k^{(m)} z^k,   \qquad \qquad  (m = 0,1,2...)
\end{align*}
where
\begin{align*}
a_0^{(m)} = \frac{1}{3^m} \qquad a_1^{(m)} = \frac{m}{3^{m-1}.4} \qquad a_2^{(m)} = \frac{m(15m+17)}{3^{m-1}.160} ...
\end{align*}
So, we can now write
\begin{align} \label{eq:integrand_approx}
e^{\alpha t }(1-t)^\alpha &= e^{-\frac{\alpha t^2}{2}} \sum_{m=0}^{\infty} \frac{(-\alpha)^m}{m!}  t^{3m} \sum_{k=0}^{\infty} a_k^{(m)}t^k = e^{-\frac{\alpha t^2}{2}} \sum_{n=0}^{\infty} \sum_{m=0}^n  \frac{(-\alpha)^m}{m!}a_{n-m}^{(m)} t^{n+2m}.
\end{align}
Substituting \eqref{eq:integrand_approx} in \eqref{eq:Iab},
\begin{align}\label{eq:Iab2}
I(\alpha, \beta) =  \sum_{n=0}^{\infty} \sum_{m=0}^n  \frac{(-\alpha)^m}{m!}a_{n-m}^{(m)} \int_0^{\beta} e^{-\frac{\alpha t^2}{2}} t^{n+2m} {\rm d}t.
\end{align}
With the substitution $t=\beta \sqrt{u}$ in the following integral, we get
\begin{align}\notag
\int_0^{\beta} e^{-\frac{\alpha t^2}{2}}t^{n+2m} {\rm d}t &= \frac{1}{2} \beta^{n+2m+1} \int_0^1 e^{-\frac{\alpha \beta^2 u}{2}} u^{\frac{n-1}{2} + m } {\rm d}u. \\
\intertext{With a change of variable, $\frac{\alpha \beta^2 u}{2} = s$, we get} \label{eq:int_step1}
\int_0^{\beta} e^{-\frac{\alpha t^2}{2}}t^{n+2m} {\rm d}t &=\frac{1}{2} \beta^{n+1+2m}  \bigg(\frac{2}{\alpha \beta^2}\bigg)^{\frac{n+1}{2}+m} \int_0^{\frac{\alpha \beta^2}{2}} e^{-s} s^{\frac{n-1}{2}+m} {\rm d}s .
\end{align}
We know by definition that 
\begin{align}\label{eq:gamma_def}
\gamma(a,x) = \int_0^x e^{-t} t^{a-1} {\rm d}t &= \Gamma(a) x^a \gamma^*(a,x).
\end{align}
Here, the fractional powers of $x$ are to be understood as having principal values.
Substituting \eqref{eq:gamma_def} in \eqref{eq:int_step1},
\begin{align}\label{eq:Iab3}
\int_0^{\beta} e^{-\frac{\alpha t^2}{2}}t^{n+2m} {\rm d}t &= \frac{1}{2} \beta^{n+1+2m} \Gamma\bigg(\frac{n+1}{2}+m\bigg) \gamma^*\bigg(\frac{n+1}{2}+m, \frac{\alpha \beta^2}{2} \bigg) \\
&= \pm \gamma\bigg(\frac{n+1}{2}+m,\frac{\alpha \beta^2}{2} \bigg) .
\end{align}
Now we define,
\begin{align}\label{eq:Fmn}
F_{m,n}(\alpha, \beta)  = \bigg(\frac{\alpha}{2}\bigg)^{\frac{n+1}{2}+m} \frac{1}{2} \beta^{n+1+2m} \Gamma\bigg(\frac{n+1}{2} + m\bigg) \gamma^*\bigg(\frac{n+1}{2}+m, \frac{\alpha \beta^2}{2} \bigg).
\end{align} 
Substituting \eqref{eq:Fmn} and \eqref{eq:Iab3} in \eqref{eq:Iab2},
\begin{align}\label{eq:Iab4}
I(\alpha, \beta) &\sim \frac{1}{2}\sum_{n=0}^\infty A_n(\alpha, \beta) \bigg(\frac{2}{\alpha} \bigg)^{\frac{n+1}{2}}, \\
\intertext{where}\label{eq:An}
A_n(\alpha, \beta) &= \sum_{m=0}^n \frac{(-2)^m}{m!} a_{n-m}^{(m)} F_{m,n}(\alpha, \beta) \\
&=
\begin{dcases}
\sum_{m=0}^n \frac{(-2)^m}{m!} a_{n-m}^{(m)} \gamma\bigg(\frac{n+1}{2}+m,\frac{\alpha \beta^2}{2} \bigg) \qquad &\beta>0\\
(-1)^{n+1} \sum_{m=0}^n \frac{(-2)^m}{m!} a_{n-m}^{(m)} \gamma\bigg(\frac{n+1}{2}+m,\frac{\alpha \beta^2}{2} \bigg) \qquad &\beta<0
\end{dcases}.
\intertext{Note that,} \notag
A_0 &= \sqrt{\pi} \erf\bigg(\sqrt{\frac{\alpha \beta^2}{2}} \bigg), 
\quad \text{where} \  \notag \erf(z) = \frac{2}{\sqrt{\pi}}\int_0^{z} e^{-t^2} {\rm d}t = \frac{1}{\sqrt{\pi}} \gamma\bigg(\frac{1}{2}, z^2\bigg), \\ \notag
A_1 &= \frac{2}{3} \bigg[ e^{-\frac{\alpha \beta^2}{2}}\Big(1+\frac{\alpha \beta^2}{2}\Big)-1\bigg].
\end{align}
Using this key intermediate result, we will now prove our main result by deriving an asymptotic estimate for the incomplete gamma function $\gamma(\alpha+1, \alpha+\xi)$ as $\alpha \to \infty$.
\begin{align} \notag
\gamma(\alpha+1, \alpha+\xi) = \int_0^{\alpha+\xi} e^{-t}t^{\alpha} {\rm d}t &\stackrel{(a)}{=} \int_{-\xi/\alpha}^1 e^{-\alpha+\alpha\tau} \big(\alpha(1-\tau)\big)^{\alpha} \alpha {\rm d} \tau \\ \label{eq:asym1}
&= e^{-\alpha}\alpha^{\alpha+1}\Big[ I(\alpha,1) - I(\alpha, -\xi/\alpha) \Big],
\end{align}
where (a) follows from the substitution $t = \alpha(1-\tau)$.
For $\beta =1$, we observe that 
\begin{align*}
\lim_{\alpha \to \infty} F_{m,n} (\alpha, \beta) = \pm \lim_{\alpha \to \infty} \gamma \bigg(\frac{n+1}{2}+m, \frac{\alpha\beta^2}{2} \bigg) = \pm \Gamma\bigg(\frac{n+1}{2}+m\bigg).
\end{align*}
For $\beta = -\xi/\alpha$, this limit results in three sub-cases depending on $\xi$ relative to $\alpha$:
\begin{align*}
\lim_{\alpha \to \infty} F_{m,n}(\alpha, \beta) = \begin{dcases}
0, \qquad \qquad & \xi = o(\sqrt{\alpha}) \\
\pm \gamma\bigg(\frac{n+1}{2}+m,k'\bigg), \qquad \qquad & lim(\xi \alpha^{-1/2}) = k \\
\pm \Gamma\bigg(\frac{n+1}{2}+m\bigg), \qquad \qquad & \sqrt{\alpha} = o(\xi) 
\end{dcases}.
\end{align*}
Using these limits in \eqref{eq:Iab4} and substituting the resulting expression in \eqref{eq:asym1}, we obtain the asymptotic expansion of incomplete gamma function for the special case.
Now, considering $\xi = \sqrt{2\alpha}y$, we get
\begin{align}
\gamma(\alpha+1,\alpha+\sqrt{2\alpha}y) \sim \sqrt{\frac{\alpha}{2}}e^{-\alpha} \alpha^{\alpha} \sum_{n=0}^\infty B_n(y)\Big(\frac{2}{\alpha}\Big)^{n/2}, \qquad \alpha \to \infty 
\end{align}
where,
\begin{align}
B_n(y) = \begin{dcases}
\sum_{m=0}^n \frac{(-2)^m}{m!}a_{n-m}^{(m)} \Gamma\bigg(\frac{n+1}{2}+m, y^2\bigg), \quad &y<0 \\
\sum_{m=0}^n \frac{(-2)^m}{m!}a_{n-m}^{(m)} \Bigg[\Gamma\bigg(\frac{n+1}{2}+m\bigg)+(-1)^m\gamma\bigg(\frac{n+1}{2}+m, y^2\bigg)\Bigg], \quad &y>0 
\end{dcases}.
\end{align}
Expanding the summation and ignoring the terms for $n\geq 2$, we obtain the same result for $y \lessgtr 0$ as follows:
\begin{align}\label{eq:asym2}
\gamma(\alpha+1,\alpha+\sqrt{2\alpha}y)  = \sqrt{\frac{\alpha}{2}} e^{-\alpha} \alpha^{\alpha} \Bigg[ \sqrt{\pi} + \sqrt{\pi}\erf\big(\sqrt{y^2}\big)-\frac{2}{3}\sqrt{\frac{2}{\alpha}}(1+y^2)e^{-y^2} + \ncalO\big(\alpha^{-1}\big) \Bigg].
\end{align}
We also know that~\cite{stirling} 
\begin{align}\label{eq:gamma_asym}
\Gamma(\alpha+1) = \sqrt{2\pi\alpha}e^{-\alpha}\alpha^{\alpha} \big[1 + \ncalO(\alpha^{-1})\big].
\end{align}
Therefore, from \eqref{eq:asym2} and \eqref{eq:gamma_asym},
\begin{align}
\frac{\Gamma(\alpha+1, \alpha+\sqrt{2\alpha}y)}{\Gamma(\alpha +1)} &= \frac{1}{2} - \frac{1}{\sqrt{\pi}}\erf(y) + \frac{1}{3}\sqrt{\frac{2}{\alpha\pi}}(1+y^2)e^{-y^2} + \ncalO\big(\alpha^{-1}\big).
\end{align}

\subsection{Proof of Lemma \ref{lem:rho}}\label{app:rho}
Since $X_i = V_i - \mu_{V_i}$ and $V_i = U_i^{-\alpha}$, the conditional PDF of $X_i$ is given by
\begin{align}\label{equ:fxi}
f_{X_i}(x_i | r, u_1, x_0) = f_{V_i}(x_i + \mu_{V_i}) = f_{U_i}\big( (x_i + \mu_{V_i})^{-1/\alpha} \big). 
\end{align} 
Now, by definition,
\begin{align}\label{equ:emodxi3}
\nbbE \big[ |X_i|^3 | R, U_1, x_0 \big] = \int_{w_p^{-\alpha}-\mu_{V_i}}^{u_1^{-\alpha} - \mu_{V_i}} |x_i|^3 f_{X_i}(x_i|r,u_1,x_0) {\rm d}x_i.
\end{align}
We know that $w_p^{-\alpha}$ is the smallest interference that could be caused by any of the transmitting nodes at the receiver. Since $w_p^{-\alpha}$ will always be smaller than the mean interference $\mu_{V_i}$, $w_p^{-\alpha} - \mu_{V_i}<0$. Therefore, we split the integral and change the limits accordingly. Substituting \eqref{equ:fxi} in \eqref{equ:emodxi3} and using the conditional distribution of $U_i$ from Lemma \ref{lem:pdfui}, we obtain the final expression.

\subsection{Proof of Theorem \ref{thm:bounds}}\label{app:bounds}
The coverage probability is given by
\begin{align}\notag
P_c &= \P(\sir > \T) = \P \bigg(\frac{R^{-\alpha}}{U_1^{-\alpha} + \sum_{i=2}^{N-1}U_i^{-\alpha}} > \T \bigg) \\ \notag
&= \bigintsss_h^{w_p} \bigintsss_r^{w_p} \P \bigg( \sum_{i=2}^{N-1}U_i^{-\alpha} < \T^{-1}R^{-\alpha} - U_1^{-\alpha} \Big| R, U_1, x_0\bigg) f_{R, U_1}(r, u_1 | x_0) {\rm d}u_1 {\rm d}r \\ \notag
&= \bigintsss_h^{w_p}  \bigintsss_r^{w_p} \P \bigg( \sum_{i=2}^{N-1}X_i < \Big(\T^{-1}R^{-\alpha} -U_1^{-\alpha} - (N-2)\mu_{V_i} \Big)\Big| R, U_1, x_0\bigg) f_{R,U_1}(r,u_1|x_0) {\rm d}u_1 {\rm d}r. 
\end{align}
Let $M_{N-2} = \frac{\sum_{i=2}^{N-1}X_i }{\sqrt{(N-2) \sigma^2}}$, where $\sigma^2 = \nbbE [X_i^2 | R, U_1, x_0]$, and $F_{M_{N-2}}(m_{N-2} | R, U_1, x_0)$ be the conditional CDF of $M_{N-2}$.
The coverage probability can now be written as
\begin{align}\label{equ:pc_int}
P_c &= \bigintsss_h^{w_p} \bigintsss_r^{w_p}  F_{M_{N-2}} \bigg( \frac{\T^{-1}r^{-\alpha} - u_1^{-\alpha}-(N-2)\mu_{V_i}}{\sqrt{(N-2) \sigma^2}} \bigg|r, u_1, x_0\bigg) f_{R, U_1}(r, u_1|x_0) {\rm d}u_1 {\rm d}r. 
\end{align}
By BET, the CDF term in the integrand is related to the CDF of standard normal distribution $\Phi(x)$ as follows:
\begin{align} \notag
&\bigg|F_{M_{N-2}} \Big( \calG\big(\T, \alpha, N,R, U_1, x_0\big) \big| R, U_1, x_0 \Big) - \Phi\Big( \calG\big(\T, \alpha, N, R,U_1, x_0\big)  \Big)  \bigg| \leq \frac{C\rho}{\sigma^3 \sqrt{N-2}} \qquad \intertext{i.e.} \notag
&\Phi \Big( \calG\big(\T, \alpha, N, R,U_1, x_0\big)  \Big) - \frac{C\rho}{\sigma^3 \sqrt{N-2}} \leq F_{M_{N-2}}\Big( \calG\big(\T, \alpha, N, R,U_1, x_0\big) \big| R,U_1, x_0 \Big)  \\ \label{equ:pc_bounds}
& \qquad \qquad \leq \Phi \Big( \calG \big(\T, \alpha, N, R, U_1, x_0\big) \Big) + \frac{C\rho}{\sigma^3 \sqrt{N-2}}, 
\end{align}%
where $\calG(\T, \alpha, N,R, U_1, x_0) = \frac{\T^{-1}R^{-\alpha} - U_1^{-\alpha} - (N-2)\mu_{V_i}}{\sqrt{(N-2) \sigma^2}}$. Substituting \eqref{equ:pc_bounds} in \eqref{equ:pc_int} and integrating the resulting inequality, we get the bounds of coverage probability.
\def\baselinestretch{1.3}
\bibliographystyle{IEEEtran}
\bibliography{journal1_v0.9}

\end{document}

%% file: notation.tex
\def\nb0{{\mathbf{0}}}
\def\nb1{{\mathbf{1}}}

\def\ncalO{{\mathcal{O}}}

\def\ncalU{{\mathcal{U}}}

\def\nbbE{{\mathbb{E}}}

\def\nbbR{{\mathbb{R}}}

\newtheorem{lemma}{Lemma}
\newtheorem{thm}{Theorem}

\newtheorem{cor}{Corollary}

\def\P{\mathbb{P}}
\def\pc{\mathtt{P_c}}

\def\V{\operatorname{Var}}

\def\erf{\operatorname{erf}}

\def\T{\beta}							

\def\sir{\mathtt{SIR}}

\def\calA{\mathcal{A}}

\def\calB{\mathcal{B}}

\def\calL{\mathcal{L}}
\def\calG{\mathcal{G}}

